\documentclass[%
 reprint,
nofootinbib,
 amsmath,amssymb,
 aps,
]{revtex4-2}

\usepackage{graphicx}
\usepackage{dcolumn}
\usepackage{bm}

\usepackage[hidelinks]{hyperref}
\usepackage{float}
\usepackage{physics}
\usepackage{tikz}
\usetikzlibrary{shapes,arrows,positioning}
\usepackage{amssymb}
\usepackage{graphicx}
\usepackage{pgf}
\usepackage{pgfplots}
\usepackage{listings}
\usepackage{rotating}
\usepackage[none]{hyphenat}
\usepackage{comment}
\usepackage{csquotes}

\begin{document}

\title{Spectrum of radiation from global strings and the relic axion density} 
 
\author{Richard A. Battye}
\email{Richard.Battye@manchester.ac.uk}
\affiliation{Jodrell Bank Centre for Astrophysics, Department of Physics and Astronomy, University of Manchester, Oxford Road, Manchester M13 9PL, United Kingdom}

\author{Lukasz P. Bunio}
\email{lukasz.bunio@postgrad.manchester.ac.uk}
\affiliation{Jodrell Bank Centre for Astrophysics, Department of Physics and Astronomy, University of Manchester, Oxford Road, Manchester M13 9PL, United Kingdom}

\author{Steven J. Cotterill}
\email{steven.cotterill@manchester.ac.uk}
\affiliation{Jodrell Bank Centre for Astrophysics, Department of Physics and Astronomy, University of Manchester, Oxford Road, Manchester M13 9PL, United Kingdom}
  
\author{Pranav B. Gangrekalve Manoj}
\email{pbg@sfu.ca}
\affiliation{Jodrell Bank Centre for Astrophysics, Department of Physics and Astronomy, University of Manchester, Oxford Road, Manchester M13 9PL, United Kingdom}

\date{\today} 

\begin{abstract}
We discuss key aspects of the nature of radiation from global strings and its impact on the relic axion density. Using a simple model we demonstrate the dependence on the spectrum of radiation emitted by strings. We then study the radiation emitted by perturbed straight strings paying particular attention to the difference between the overall phase of the field and the small perturbations about the string solution which are the axions. We find that a significant correction is required to be sure that one is analyzing the axions and not the self-field of the string. Typically this requires one to excise a sizeable region around the string - something which is not usually done in the case of numerical field theory simulations of string networks. We have measured the spectrum of radiation from these strings and find that it is compatible with an exponential, as predicted by the Nambu-like Kalb-Ramond action, and in particular is not a ``hard'' spectrum often found in string network simulations. We conclude by attempting to assess the uncertainties on relic density and find that this leads to a wide range of possible axion masses when compared to the measured density from the Cosmic Microwave Background, albeit that they are typically higher than what is predicted by the Initial Misalignment Mechanism. If the primary mode of decay is via a ``soft spectrum'' from loops produced close to the backreaction scale we find that $m_{\rm a}\approx 160\,\mu{\rm eV}$ and a detection frequency $f\approx 38\,{\rm GHz}$. If axions are primarily emitted directly by the string network, and we use emission spectra reported in typical field theory simulations, then $m_{\rm a}\approx 4\,\mu{\rm eV}$ and $f\approx 1\,{\rm GHz}$, however this increases to $m_a \approx 125\,\mu{\rm eV}$ and $f\approx 30\,{\rm GHz}$ using our spectra for the case of an oscillating string. In all scenarios there are significant remaining uncertainties that we delineate.
\end{abstract}

\maketitle

\pagenumbering{arabic}

\section{\label{sec:level1}Introduction}

The axion is currently the most studied of all the possible dark matter candidates~\cite{marsh2016axion}. It results from imposition of the Peccei-Quinn (PQ) symmetry~\cite{peccei1977constraints} that can ameliorate the strong CP problem. The pseudo-Nambu Goldstone Boson associated with the symmetry~\cite{ref:weinberg,ref:wil} can be incorporated within the framework of the Standard Model of Particle Physics, the most popular variants being due to Dine-Fischler-Srednicki-Zhitnitsky (DFSZ)~\cite{ref:DFSZ,ref:Zhit} and Kim-Shifman-Vainshtein-Zakharov (KSVZ)~\cite{ref:K,ref:SVZ}. 
 
The dominant mechanism for the production of axions is typically not thought to be via the standard thermal scenario~\cite{Turner:1986tb}, but rather through either the ``Initial Misalignment Mechanism'' (IMM)~\cite{ref:misalign1,ref:misalign2,ref:misalign3}, or their production by the topological defects which form during the PQ phase transition~\cite{Sikivie:1982qv,Davis:1986xc}. The IMM is often connected to scenarios where the PQ phase transition takes place before inflation, whereas the production by topological defects is usually assumed to dominate when the phase transition happens after inflation. 

Typically the PQ phase transition is believed to take place at an intermediate scale while many, but not all, models of inflation involve a higher energy scale, often connected to a Grand Unified Theory (GUT) scale. Hence, understanding the production of axions by topological defects is of critical importance.

The relic density in the IMM scenario can be calculated by solving the ordinary differential equation for the axion density coupled to the expansion history of the Universe. It also requires knowledge of the evolution of the axion mass which is zero initially, but becomes massive during the QCD phase transition~\cite{ref:WS,ref:BaeMisalign,ref:WS,Borsanyi:2015cka,Borsanyi:2016ksw,PhysRevD.85.054503,PhysRevD.106.074501,Bonati:2015vqz,PETRECZKY2016498}.

Scenarios involving topological defects are much more difficult to model. This is because they involve solving partial differential equations covering a wide range of scales to encompass the core of the defects and the size of the Universe, albeit in the radiation-dominated era, over a substantial fraction of the Universe's history; approximations are necessary to do this and, hence, the predictions of the axion mass are much less certain.

There are two phases in scenarios where topological defects are formed: global cosmic strings are formed at the PQ phase transition where the axions are effectively massless and then domain walls form, connected to the strings, around the QCD phase transition when the axion becomes massive. If these domain walls do not decay then there will be a domain wall problem, making such a scenario not viable. However, it is possible that the hybrid string-wall network can decay, allowing relaxation to the CP conserving value of the $\theta$-angle. In this article we will largely be concerned with the production of axions from global strings, but we will also comment on those that would be formed in the final period of their decay from the network in section~\ref{sec:discussion}.

The predicted relic abundance of axions produced by global cosmic strings is controversial and has been discussed by a number of authors. Historically, two scenarios, denoted A and B below, have been suggested.
\begin{itemize}
    \item {\it Scenario A: }It was suggested in refs.~\cite{Davis:1986xc,Davis:1989nj,battye1994global,Battye:1994au,battye1997recent} that the power of axions emitted by a global string would peak with a frequency close to the fundamental mode of the string perturbation. This is predicted~\cite{Vilenkin:1986ku} by the Kalb-Ramond action~\cite{kalb1974classical} which treats the string as Nambu-like~\cite{nambu1974strings}, but coupled to an anti-symmetric tensor field~\cite{Davis:1988rw}.
    \item {\it Scenario B: }An alternative suggestion was made in refs.~\cite{harari1987evolution} and has been claimed to be supported by numerical simulations~\cite{Hagmann:1990tj}. In this scenario, the string emits axions with a hard spectrum, which has a logarithmic divergence, meaning that it emits significantly at all frequencies from the fundamental mode to the core width.
\end{itemize}
Scenario A typically leads to a larger relic density than B. The precise details will be reviewed in sections~\ref{sec:delineaate} and \ref{sec:estimates}, and - when compared to the Cold Dark Matter (CDM) density, $\Omega_{\rm c}h^2\approx 0.12$ inferred by observation of the Cosmic Microwave Background (CMB)~\cite{Planck:2018vyg} -  gives a somewhat larger value of the axion mass, $m_{\rm a}$, and a lower value of the axion decay constant, $f_{\rm a}\propto m_{\rm a}^{-1}$~\cite{Battye:1994au,battye1997recent}; however, the uncertainties are large.

This historical work typically only considered specific solutions such as perturbed straight strings and circular loops. Based on the quantitative agreement between the predictions of Kalb-Ramond action and field theory simulations~\cite{battye1994global}, it was suggested that the evolution of global strings would be similar to that found in ``Nambu'' string simulations, most importantly in defining the density of strings in the scaling solution. This includes the production of loops which it was argued would be the most significant source of axions~\cite{Battye:1994au,battye1997recent}.

Numerous field theory simulations of strings have been performed over the years of both global $U(1)$ strings~\cite{Yamaguchi:1999dy,Kawasaki:2018bzv,saikawa2024spectrum,Kaltschmidt:2025nkz,buschmann2020early,buschmann2022dark,Benabou:2024msj,hindmarsh2020scaling,Klaer:2017ond,Klaer:2019fxc,Hindmarsh:2021vih, Kim:2024wku, PhysRevD.83.123531, Correia:2024cpk, Gorghetto:2020qws, kim2024scalingscosmicstrings,Gorghetto:2018myk,Correia:2025nns} and those in the Abelian-Higgs model~\cite{Hindmarsh:2017qff, PhysRevD.65.023503}. A general conclusion of this large body of work is that the production and evolution of loops does not happen in a way that is compatible with simulations based on evolving the Nambu action~\cite{Hindmarsh:2008dw,Hindmarsh:2021mnl,Baeza_Ballesteros_2025}. In particular, loops are observed to decay in time shorter than one oscillation. Moreover the density of strings measured in field theory is much smaller than predicted by the ``Nambu'' simulations~\cite{Allen:1990tv,Shellard:1989yi,Bennett:1987vf,Bennett:1989ak,Bennett:1989yp}. Although this work is not entirely compatible with scenario B above, the spectrum of radiation has been claimed to be ``hard'' and to be close to the regime where it is significantly influenced by the scale of the core of the string rather than the size of the perturbations on the string. This would suggest that the network simulations are not compatible with perturbed straight string simulations of ref.~\cite{battye1994global,battye1997recent} and by implication the Kalb-Ramond action. Why this might be the case is an open question.

Even the largest simulations performed~\cite{Benabou:2024msj} do not have the dynamic range, defined by the ratio of the string correlation length, $\Delta\sim t$ to core width, $\delta$, to fully match what is necessary to make predictions of the relic axion density. This is because one has to resolve both scales in the simulation, requiring $\log(\Delta/\delta)\approx 70$. The very largest simulations use {\em Adaptive Mesh Refinement} (AMR)~\cite{drew2022radiation,Drew:2022iqz,buschmann2022dark} to achieve $\log(\Delta/\delta)\approx 9$. This is impressive, but is still significantly lower than the value relevant to the real Universe, and hence one is forced to make an extrapolation (usually based on some analytic model) over many orders of magnitude, and regimes where different phenomena dominate the dynamics. This could play some role in the discrepancy described above.

In addition to simulations from random initial conditions that model the evolution of a network of strings, attempts have been made to construct loops of strings~\cite{Saurabh:2020pqe, Baeza_Ballesteros_2024}. It was suggested that the loops emitted a hard spectrum with $q\approx 1$, although the loops did not last for a significant number of oscillations. A circular loop was studied in ref.~\cite{battye1994global}. This decays in a single oscillation time and will emit a hard spectrum. An important question to answer is whether loops produced in the way suggested in ref.~\cite{Saurabh:2020pqe} and indeed those produced in network simulations are just too small to last long enough for them to be sufficiently different from a circular loop.

In this paper we will first review the key aspects of the calculation of the relic axion density with respect to a model. This is an adaptation of that used in refs.~\cite{Battye:1994au,battye1994global} - which is in turn based on the one-scale model treatment in ref.~\cite{vilenkin1994cosmic}. The calculation is improved by including direct production of axions by the network and what becomes clear is that, whether loops or the long strings are the dominant mechanism, the spectrum of radiation is the key uncertainty. Subsequently, we show that the quantity typically used to calculate the axion density in numerical field theory simulations is dominated by the self-field of the string and that some restriction on the lattice sites included in the calculation is required to be sure that one is measuring the spectrum of emitted axions. This is relatively easy to do in the case of a perturbed straight string, but is more difficult in the context of a network simulation - the main reason being the strong coupling of the axion field to the string self-field. On the basis of this,  we attempt to quantify the uncertainties in calculating the relic axion density based on a parametric model, ultimately concluding that the range of axion masses predicted from global string radiation is somewhat larger than that from the IMM.

\section{Delineating the relic axion density from strings}
\label{sec:delineaate}
In this section we will explain how the axion density depends on cosmology, the various parameters associated with the evolution of the string network and most importantly the spectrum of radiation emitted by the strings. We will consider two components: axions emitted by string loops which are themselves created by the long string network that itself also emits axions. We will assume that a fraction ${\cal F}_{\ell}$ of the energy required to maintain scaling of the long string network is emitted into loops which eventually decay into axions and a fraction $1-{
\cal F}_\ell$ is emitted directly into axions by the long string network. In what follows we will derive expressions for the the relic axion density assuming that the axion number density to entropy ratio remains constant after temperature $T_{\rm AD}$, which is around the time of the QCD phase transition. In section~\ref{sec:estimates} we will use these estimates to discuss the uncertainties on the mass of the axion assuming that it provides the cold dark matter density measured to be $\Omega_{\rm c}h^2\approx 0.12$ using CMB measurements by the {\it Planck} satellite~\cite{Planck:2018vyg}. 

Before considering the scenario where the axions are emitted from both loops and long strings we will consider each of the components separately. The long string density is 
\begin{equation}
\rho_\infty={\zeta\mu\over t^2}\,,
\end{equation}
where $\zeta$ is a constant quantifying the number of long strings per horizon volume and $\mu\approx\pi f_{\rm a}^2\log(\Delta/\delta)$ with $\delta\sim f_{\rm a}^{-1}$ denoting the width of the string. Simulations of Nambu strings find that $\zeta\approx 13$ whereas those for numerical field theory simulations estimate $\zeta$ in the range $0.5-1.5$ and, therefore, we will consider $\zeta\sim 1$ to be the fiducial value for that scenario. This must satisfy the equation of motion
\begin{equation}
{\dot\rho}_\infty+2H(1+\langle v^2\rangle)\rho_\infty=-{c\rho_\infty\over L}
\end{equation}
where $L=\zeta^{-1/2}t$, $\langle v^2\rangle$ is the r.m.s. string velocity and $c$ is known as the ``chopping efficiency''. Following the discussion in chapter 9 of ref.~\cite{vilenkin1994cosmic}, if we ignore the logarithmic time dependence of $\mu$, we find that 
\begin{equation}
    c=\zeta^{-1/2}(1-\langle v^2\rangle)\,.
\end{equation}
This is an important quantity since it normalizes the amounts of energy that can be emitted either directly into axions, or into loops. 

\begin{figure*}[!t]
\vspace{-3.5\baselineskip} 

        \includegraphics[scale=0.5]{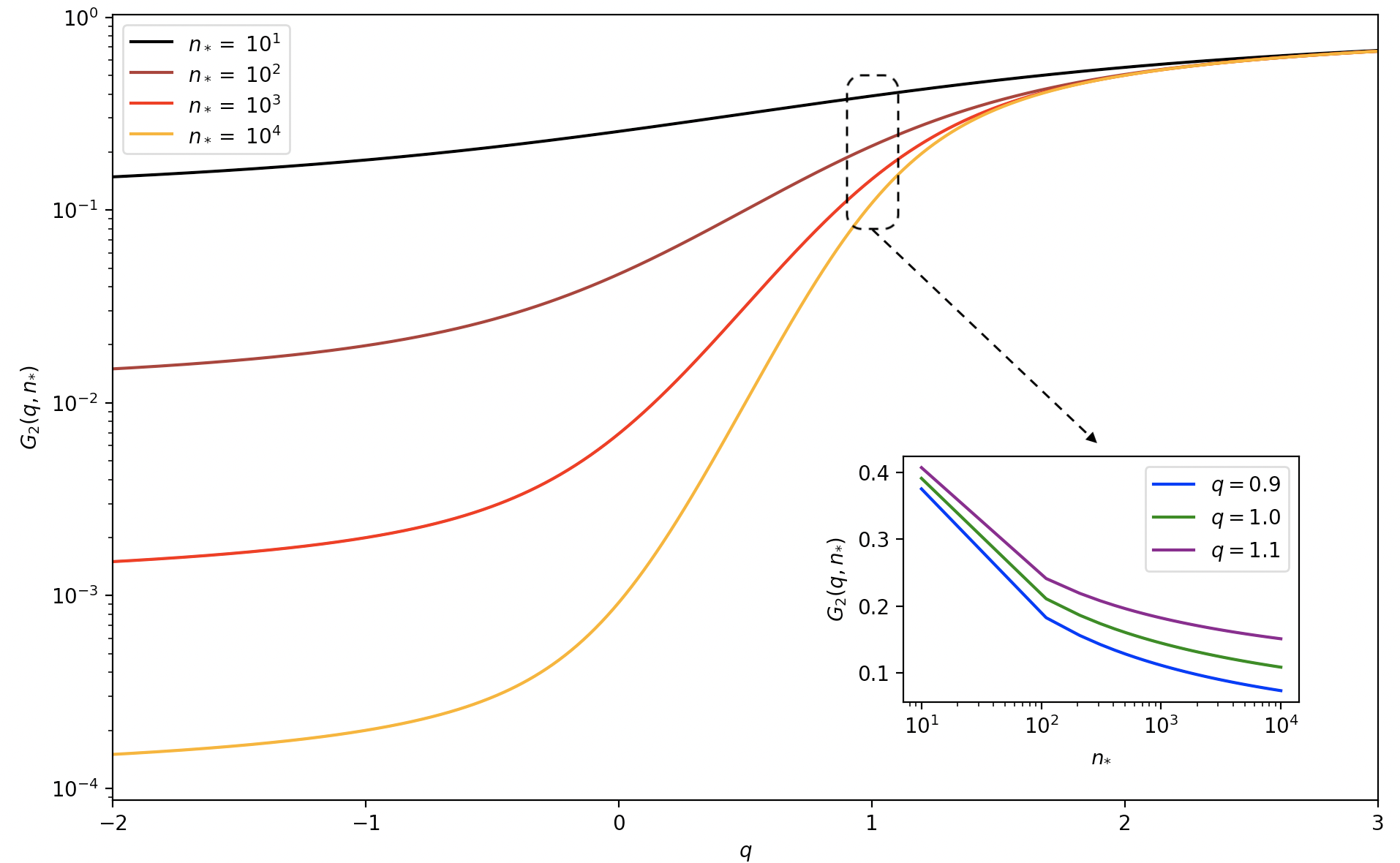}
    
\vspace{-1\baselineskip} 
\caption{The function $G_2(q,n*)$ as a function of $q$ for various values of $n_*$. The clear features are an asymptote for $q\rightarrow\infty$ and a strong dependence on $n_*$ for large negative $q$. We have also included as an inset the dependence on $n_*$ for values of $q$ close to one. Remembering that the axion mass at fixed $\Omega_{\rm a}h^2$ is proportional to $G_2$, this makes it clear that the predicted value depends very strongly on the spectrum unless $q\gg 1$. Note also that the contributions from both loop and long strings will have a similar functional form, but possibly very different parameters.}
\label{g2}
\end{figure*}

In order to calculate the relic density of axions, we first calculate the axion number density to entropy ratio
\begin{equation} 
{n_a\over s}={\int_{4\pi\over\alpha t}^{\infty}d\omega {1\over\omega}{d\rho_a\over d\omega}\over {2\pi^2\over 45}{\cal N}_{\rm S}T^3}\,,
\end{equation}
where ${\cal N}_S$ is the entropy weighted number of degrees of freedom and $\textstyle{\partial\rho_{\rm a}\over\partial\omega}$ is the spectral density of massless axions emitted. The number density to entropy ratio is constant from some epoch, defined by $T=T_{\rm AD}$ close to the QCD phase transition to the present day. Using this we deduce that the density parameter for the axions is given by 
\begin{equation}
\Omega_{\rm a}={32\pi^2\sqrt{\pi}\over 3\sqrt{45}} {{\cal N}_{S0}m_{\rm a}T_{\rm CMB}^3\over m_{\rm pl}^3H_0^2{\cal N}_S{\cal N}^{-1/2}}{n_{\rm a}t_{\rm AD}\over T_{\rm AD}}\,,
\end{equation}
where $n_{\rm a}$ is the axion number density at $T_{\rm AD}$ and $t_{\rm AD}$ is the corresponding time. $H_0=100h\,{\rm km}\,{\rm sec}^{-1}\,{\rm Mpc}^{-1}$ is the Hubble constant at the present day, $T_{\rm CMB}=2.725\,{\rm K}$ and ${\cal N}$ is the number of degrees of freedom at the point when the evolution becomes adiabatic. ${\cal N}_{S0}=43/11\approx 3.91$ is the entropy weighted number of relativistic degrees of freedom for 3 species of neutrinos at the present day.

\subsection{Axions from string loops}

First, let us assume that all the energy is emitted into loops which was the assumption made in refs.~\cite{Battye:1994au,battye1997recent}.
We will consider loops produced at time $t_i$ to have size $\ell_i=\alpha t_i$. The power emitted by these loops is independent of the loop length and is given by $P=\kappa\mu=\Gamma_{\rm a}f_{\rm a}^2$. The lifetime of these loops is $(\alpha/\kappa)t_i$. From \cite{Battye:1994au,battye1997recent} the spectral density of axions emitted by the distribution of loops is given by 
\begin{eqnarray}
{d\rho_{\rm a}\over d\omega}(t)={4\Gamma_{\rm a}f_a^2\nu\over 3\omega\kappa^{3/2}t^2}\int_0^{\alpha\omega t}dz g(z)\cr\times \left[\left(1+{z\over \omega\kappa t}\right)^{-3/2}-\left(1+{\alpha\over\kappa}\right)^{-3/2}\right]\,
\label{drhoa}
\end{eqnarray}
for $\omega>4\pi/(\alpha t)$ where $\nu=\beta\zeta(1-\langle v^2\rangle)\alpha^{1/2}\left(1+{\kappa\over\alpha}\right)^{3/2}$ and the function $g(z)$, which encodes the spectrum of radiation emitted by the strings in normalized so that 
\begin{equation}
\int_0^{\infty}dz\,g(z)=1\,.
\label{norm}
\end{equation}
The parameter $\beta$ takes into account that the loops are not produced at rest, but with a relativistic centre-of-mass  velocity, and hence some of the energy will be dissipated by the redshifting of this velocity. 
Typical values measured from simulations using the Nambu action are $\beta\approx 1/\sqrt{2}$, $\zeta\approx 13$, $\langle v^2\rangle\approx 0.35$ and $\Gamma_{\rm a}\approx 65$. Note that in refs.~\cite{Battye:1994au,battye1997recent} (i) the normalization was defined slightly differently - being $=\Gamma_{\rm a}$ - but this factor has been included in (\ref{drhoa}) instead. (ii) The factor $\beta$ was called $g$ in previous work.

Here, we will consider two cases. The first is a simple power-law  function 
\begin{equation}
g(z)=A\theta(z-4\pi)\theta(4\pi n_*-z)z^{-q}\,,
\end{equation}
where $n_*>1$. In this case the normalization yields
\begin{equation}
A=(q-1)(4\pi)^{q-1}\left[1-{1\over n_*^{q-1}}\right]^{-1}\,,
\end{equation}
if $q\ne 1$. In the case of $q>1$ we can take the limit $n_*\rightarrow\infty$, but not when $q\le 1$. If $q=1$ then $A=\Gamma_a/\log n_*$. The Second case we will consider is an exponential fall off with
\begin{equation}
g(z)=A\theta(z-4\pi)\exp\left[-{rz\over 4\pi}\right]\,,
\end{equation}
for which $A=r\exp[r]/(4\pi)$.

In previous work~\cite{Battye:1994au,battye1997recent} it was argued that the an approximation can be made
\begin{equation}
{d\rho_{\rm a}\over d\omega}(t)={4f_a^2\nu\Gamma_a\over 3\omega\kappa^{3/2}t^2}\left[1-\left(1+{\alpha\over\kappa}\right)^{-3/2}\right]\,,
\label{olddrho}
\end{equation} 
which leads to an axion density that is independent of the spectrum, that is, the parameters $q$ and $n_*$. As we will see this approximation is a good one if $q\gg 1$, but not for more general values of $q$ and $n_*$. There is an upper limit on the value of $n_{*}\approx \Delta/\delta\approx e^{70}$ which is very large and effectively infinite. However, one might expect the effects of radiation backreaction to reduce this somewhat~\cite{battye1994global}. 

It turns out that it is not necessary to make this approximation and the integrals needed to calculate the overall axion density can be evaluated exactly. The spectral density can be calculated from 
\begin{equation}
{\partial\rho_{\rm a}\over\partial\omega}(t)=\Gamma_{\rm a}f_{\rm a}^2\int_{t_{*}}^tdt^{\prime}\left(\textstyle{a(t^\prime)\over a(t)}\right)^3\int_0^{\alpha t^\prime}\ell d\ell \,n(\ell,t^\prime)g\left(\textstyle{a(t)\over a(t^\prime)}\omega\ell\right)\,,
\end{equation}
where $n(\ell,t)=\nu t^{-3/2}(\ell+\kappa t)^{-5/2}$ is the number density of loops of length $\ell$ and $t_*$ is the time when relativistic motion of the strings starts, that is, when the network starts to emit axions. One can take the integral over $\omega$ through the integrals for $t^{\prime}$ and $\ell$ when calculating $n_{\rm a}$ to give a factor of 
\begin{equation}
\int_0^{\infty}{dz\over z}g(z)\,,
\end{equation}
which will come up a number of times in the subsequent discussion. Using this we can deduce an expression for the density parameter
\begin{equation}
\Omega_a=N{_{\rm S}\hat\Gamma}_{\rm a}G_1\left({\alpha/\kappa}\right)G_2(q,n_*)\,.
\end{equation}
where ${\hat\Gamma}_{\rm a}=\beta\Gamma_{\rm a}\approx 46$ and the normalization factor, $N_{\rm S}$ is given by 
\begin{equation}
    N_{\rm S}={8\pi\sqrt{\pi}\over 3\sqrt{45} }{{\cal N}_{S0}\zeta(1-\langle v^2\rangle)\over {\cal N}_{\rm S}{\cal N}^{-1/2}}{m_af_a^2T_{\rm CMB}^3\over H_0^2T_{\rm AD}m_{\rm pl}^3}\,,
    \label{normalization}
\end{equation}
where we have assumed that $t_{\rm AD}\gg t_*$. Note that this normalization factor is $\propto m_{\rm a}^{-1}$ and hence high values of $m_{\rm a}$ lead to lower values of $\Omega_{\rm a}$,

This expression is the product of a combination of fundamental constants and parameters describing the cosmological evolution around the QCD phase transition, along two functions $G_1(x)$ and $G_2(q,n_*)$. The first of these is given by 
\begin{equation}
G_1(x)={8\over 3x}\left[(1+x)^{3/2}-1-{3\over 2}x\right]\,,
\end{equation}
and it describes the effects of the loop decay. For $x\ll 1$ we see that $G_1(x)\approx x$ and for $x\gg 1$ we have that $G_1(x)\approx  8\sqrt{x}/3$. In particular we see that the contribution from loops becomes unimportant if $\alpha/\kappa\rightarrow 0$. We note that, within the context of the calculation, this limit is not the same as all the energy being emitted from the long strings due to influence of the spectrum - if $\alpha\rightarrow\infty$ all the axions would be emitted with very high frequencies, reducing $n_{\rm a}$ for the same amount of energy emitted.

If we had used (\ref{olddrho}) we would have deduced that $G_1(x)\propto\left(1+x\right)^{3/2}-1$ which has the same dependence on $x$ as $x\rightarrow 0$, but it is very different for large $x$. Moreover, there would have been no dependence on the spectrum, that is, $G_2=1$. Hence, we have found an improved treatment of the contribution to the axion density from string loops.
 
The second function is defined as 
\begin{equation}
G_2=4\pi{\int_0^\infty{dz\over z}\,g(z)\over \int_0^\infty dz\,g(z)}
\end{equation}
and this encodes the effects of the radiation spectrum which is the main topic of this paper. For the power law function this is given
\begin{equation}
G_2(q,n_*)=\left(1-{1\over q}\right){1-n_*^{-q}\over 1- n_*^{1-q}}\,.
\end{equation}
This function has some interesting properties: prima-facie it might look like this function is singular at $q=0$ and goes to zero at $q=1$. However, these features are removed by realizing that 
\begin{equation}
1-n_*^{-q}=q\log n_*+{\cal O}(q^2)\,,
\end{equation}
and 
\begin{equation}
1-n_*^{1-q}=(q-1)\log n_*+{\cal O}[(q-1)^2]\,.
\end{equation}
Therefore, we see that, for example, $G_2(1,n_*)=(1-n_*^{-1})/\log n_*$. In addition, it is clear that $G_2(q,n_*)\rightarrow 1$ as $q\rightarrow \infty$ for any value of $n_*\gg 1$. 

For the exponential function 
\begin{equation}
G_2(r)=r\exp\left[r\right]E_1(r)=\int_0^\infty{e^{-t}\over 1+t/r}dt\,,
\end{equation}
where 
\begin{equation}
E_1(x)=\int_x^{\infty}{e^{-y}\over y}dy\,, 
\end{equation}
is the exponential integral which has the property that $E_1(x)\sim\exp[-x]/x$ as $x\rightarrow\infty$ and, hence, $G_2\sim 1$ for large $r$. 

\subsection{Axions from the long string network}
\label{sec:ax_long}

Now let us assume that the axions are produced directly from the string network with a spectrum which can modeled in a similar way to that from loops, albeit with possibly different parameters governing the spectrum of radiation; we will use $p$ and $m_*$ to denote fall off the spectrum and the cut-off, respectively. The spectral density is given by
\begin{equation}
    {d\rho_{\rm a}\over d\omega}(t)=\int_{t_*}^{t} dt^{\prime}\left(\textstyle{a(t^{\prime})\over a(t)}\right)^3{\mu\ell\zeta(1-\langle v^2\rangle)\over t^{\prime 3}}g\left(\textstyle{a(t^\prime)\over a(t)}\omega\ell\right)
\end{equation}
where $a(t)\propto t^{1/2}$, $\ell=\gamma t $ is a length scale defining the scale of the radiation from the network and again $g(z)$ is normalized as in (\ref{norm}). From this we can deduce that the number density of axions produced is 
\begin{equation}
    n_{\rm a}=\int_0^\infty{dz\over z}g(z)\int_{t_*}^{t}dt^{\prime}\left(\textstyle{a(t)\over a(t^\prime)}\right)^3{\mu\gamma\zeta(1-\langle v^2\rangle)\over t^{\prime 2}}\,.
\end{equation}
Note that there is an assumption, which was also the case for the loops, that the only dependence of the spectrum on $t^{\prime}$ is due to scale factor and $\ell$, that is, we have assumed that the function $g(z)$ is independent of time.

If we now make the assumption that $g(z)=B\theta(z-2\pi)\theta(2\pi m_*-z)z^{-p}$ then 
\begin{equation}
\Omega_{\rm a}=N_{\rm S}{\tilde\Gamma}_{\rm a}G_2(p,m_*)\,,
\end{equation}
where the normalization factor is the same the case of loops (\ref{normalization}) and ${\tilde\Gamma}_{\rm a}=4\gamma\mu/f_a^2\approx 4\pi\gamma\log(\Delta/\delta)$. The first constant is  the same as in the case where the energy from the long string network is first emitted into loops and then to axions via the decay of the loops, and most importantly the spectral function $G_2(p,m_*)$ is the same, albeit with possibly different parameters. This would be the case if the emission is a power law. Hence, we see that in both cases the dependence on the emission spectrum is given by the function $G_2$.

One possible natural choice is to assume that $\gamma\approx\zeta^{-1/2}$ with $\log(\Delta/\delta)\approx 70$ and hence ${\tilde\Gamma}_{\rm a}\approx 280\pi\zeta^{-1/2}$. If $\zeta\approx 13$ as one might expect for strings that are modeled with the Nambu action then  ${\tilde\Gamma}_{\rm a}\approx 240$. Whereas if $\zeta\approx 1$ as in the numerical field theory simulations, then ${\tilde\Gamma}_{\rm a}\approx 880 $. In what follows we will quantify the strength of long string radiation in terms of ${\tilde\Gamma}_{\rm a}$ and assume that it is in the range 200-1000 in order to take into account the possible uncertainties.

\subsection{Model with both loop and long string emission}

Taking the results from the last two subsections we can deduce that $\Omega_{\rm a}=N_{\rm S}M({\cal F}_\ell,{\hat\Gamma}_{\rm a},{\tilde\Gamma}_{\rm a},\alpha/\kappa,q,n_*,p,m_*)$ and 
\begin{eqnarray}
M&=&{\cal F}_\ell{\hat\Gamma}_{\rm a} G_{1}\left({\alpha/\kappa}\right)G_2(q,n_*)\cr&+&(1-{\cal F}_\ell){\tilde\Gamma}_{\rm a} G_2(p,m_*)\,,
\label{overall}
\end{eqnarray}
where as previously stated $0\le{\cal F}_\ell\le 1$ is the fraction of the energy required to maintain scaling which is emitted into loops. Just as a reminder the Nambu string simulations suggest ${\cal F}_\ell\sim 1$, whereas the numerical field theory simulations suggest ${\cal F}_\ell\ll 1$ and loop production can be ignored. In section~\ref{sec:estimates} we will quantify this estimate of the axion density in various scenarios. 

In Fig.~\ref{g2} we have plotted the function $G_2(q,n_*)$ as a function of $q$ for various values of $n_*$. Note that we will see that the axion mass will be proportional to $G_2$. As expected the function asymptotes for $q$ large and is almost insensitive to $n_*$ for $q>2$. In the key range predicted by loop configurations within the framework of the KR action $1<q<2$ we see that there is relatively strong dependence on $q$ and $n_*$ indicating that prediction for the axion mass is very sensitive to the spectrum. For $q\le 1$ there is a very strong dependence on $n_*$ since in these cases the high frequency radiation associated with the core width dominates, whereas for $q>1$ one is dominated by the low frequency modes which are related to the wavelength of the perturbation on the string for long periodic configurations, or the size of the loop. From this we can conclude that knowledge of the spectrum is a key piece of information which we need to understand. Scenario A for the radiation corresponds to $q>1$, whereas scenario B corresponds to $q\approx 1$. Some of the numerical field theory simulations of string networks have found that $q<1$ and some $q\approx 1$. Without extrapolation in $\log(\Delta/\delta)$ none have found that $q>1$, but this could be due to a lack of dynamics range.

The relative contribution from loops and long strings is given by 
\begin{equation}
    {\cal R}_L={{\cal F}_\ell\over 1-{\cal F}_\ell}{\hat\Gamma_{\rm a}\over {\tilde\Gamma}_{\rm a}}{G_1\left({\alpha/\kappa}\right)G_2(q,n_*)\over G_2(p,m_*)}\,.
\end{equation}
In the discussion of sec~\ref{sec:ax_long} we have argued that ${\hat\Gamma}_{\rm a}/{\tilde\Gamma}_{\rm a}$ in the range of $\approx 0.04-0.2$ which is a significant uncertainty. One might think that the spectrum of radiation from loops and long strings is very similar so that $q\approx p$ and $n_*\approx m_*$ and hence the factors of $G_2$ might cancel. But there there is additional uncertainty coming from ${\cal F}_\ell$ and $G_1(\alpha/\kappa)$: arguments based on the Kalb-Ramond action suggest that ${\cal F}_\ell\approx 1$ whereas numerical field theory simulations suggest ${\cal F}_\ell\approx 0$ and $\alpha/\kappa$ is also very small. It is clear that there is no consensus on the value of this ratio. Hence, in section~\ref{sec:estimates} we will consider each of the possibilities when attempting to predict the axion mass.

In the IMM scenario, the axion number density at $T_{\rm AD}$ is given by 
\begin{equation}
    n_{\rm a}={1\over 2}m_{\rm a}(T_{\rm AD})f_{\rm a}^2\langle\theta_{\rm i}^2\rangle\,,
\end{equation}
where $m(T_{\rm AD})$ is the axion mass at the time when the adiabatic evolution starts defined by $m_a(T_{\rm AD})=3H$ and $\langle\theta_{\rm i}^2\rangle^{1\over 2}$ is the r.m.s. of the initial misalignment angle. Using this we can deduce that 
\begin{equation}
    \Omega_{\rm a}=N_{\rm I}\langle\theta_{\rm i}^2\rangle\,,
\end{equation}
where $N_{\rm I}=3\pi N_{\rm S}/[\zeta(1-\langle v^2\rangle)]$.

The ratio of the possible contribution from the initial misalignment mechanism to that from strings is independent of the cosmology dependent factors and the details of the QCD phase transition~\cite{PhysRevD.85.054503}~\footnote{Although it is possible for there to be an initial misalignment contribution when strings form due to a constant offset in the phase, the conventional wisdom is that the spatial variation of the phases will be set by the dynamics of the phase transition and the subsequent scaling dynamics. Since the ratio of the time of the QCD phase transition and that of PQ phase transition is very large, it is reasonable to believe that any memory of any initial misalignment would be lost.}. It is given by 
\begin{equation}
    {\cal R}_M={\Omega_{\rm a}^{\rm strings}\over\Omega_{\rm a}^{\rm IMM}}={\zeta(1-\langle v^2\rangle)\over 3\pi\langle\theta_i^2\rangle}M\,,
\end{equation}
and we will discuss the regions of parameter space where ${\cal R}_M>$ and $<1$ in section~\ref{sec:estimates}.

The value of this ratio clearly depends on the initial misalignment angle. A common choice~\cite{Turner:1985si} is $\langle\theta_{\rm i}^2\rangle=\pi^2/3$ which is based on a uniform distribution in space. We will use this as our fiducial value in what follows. If the PQ symmetry is not restored after the end of inflation, that is, the reheat temperature, $T_{\rm reh}<f_{\rm a}$, then the value of $\theta_i$ is set by that in the inflating Hubble patch. In that case, there is no obvious physics that picks out any particular value leading to no real prediction for the value of $\theta_{\rm i}$ and hence some have resorted to anthropic arguments based on the observed dark matter density. The veracity of such ideas is a matter of taste, but one could argue that the string scenario is more attractive in that it is deterministic, albeit with significant uncertainties associated with actual making a prediction since the nature of the evolution of the network is still under significant debate.

Another contribution that should be considered is axion production from hybrid string-domain wall networks. Near the QCD phase transition, that is, close to $T_{\rm AD}$, the axion acquires a mass, domain walls form and the network must decay away - if it does not then there will be a domain wall problem which restricts the number of CP conserving vacua to one. If we assume that this decay takes place then the last part of the evolution of the network could be very different to what we have modelled so far. 

If we assume that the walls form at the time when axions begin to evolve adiabatically, $t_{\rm AD}$, then the amount of energy density stored in the network is $\rho_{\infty}=\zeta\mu/t_{\rm AD}^2$. Following the calculations of the spectrum from strings and also assuming that this is almost instantaneously transformed into axions with a spectrum $\propto g_{\rm dw}(\ell\omega)$ where $\ell=\gamma_{\rm dw}t_{\rm AD}$ then we can calculate the contribution to the relic axion density. This is given by  
\begin{equation}
\Omega_{\rm a}=N_{\rm D}{\hat\Gamma}_{\rm a,dw}G_2(q_{\rm dw},n_{*\rm dw})\,,
\end{equation}
where $N_{D}=N_{\rm S}/(1-\langle v^2\rangle)$, $q_{\rm dw}$ and $n_{*\rm dw}$ characterize the spectrum of radiation emitted by the domain wall-string hybrid network and ${\hat\Gamma}_{\rm a,dw}=2\gamma_{\rm dw}\mu/f_{\rm a}^2=2\pi\gamma_{\rm dw}\log(\Delta/\delta)$. Hence, we can estimate the ratio
\begin{equation}
{\cal R}_{D}={\Omega_{\rm a}^{\rm dw}\over\Omega_{\rm a}^{\rm str}}={\hat\Gamma_{\rm a,dw}G_2(q_{\rm dw},n_{*\rm dw})\over (1-\langle v^2\rangle)M}\,.
\end{equation}
It is clear that this is typically ${\cal O}(1)$ but that the domain walls could dominate over the strings in certain regions of parameter space. This motivates investigations of the hybrid system but we will not consider them further in this work.

\section{Numerical simulations of perturbed straight strings}
\label{sec:straight}
In the previous section we have demonstrated that the relic axion density is strongly dependent on the spectrum of radiation emitted by the string network, irrespective of whether it is from loops or long strings. Here, we will perform numerical simulations of perturbed periodic straight string solutions along the lines of refs.~\cite{Davis:1989nj,battye1994global}. The key issue we want to highlight is the very significant difference between the phase of the field and the axion field that has been emitted by the strings.

 Unlike gravitational radiation from strings~\cite{battye1997gravitationalwavescosmicstrings,Sousa_2014, Datta:2025vyu, Ghoshal:2025iil} which can be treated perturbatively - one can evolve the field equations for a string network and then calculate the spectrum of gravitational radiation - the axion field is strongly coupled to that of the string and some care is needed in making the distinction. The approach of using the KR action for describing the strings is designed to ameliorate this concern, but it does also rely on treating the strings a line-like objects. Here, we will attempt to discuss this issue in the context of field theory simulations. 

The Lagrangian for the $U(1)$ Goldstone model is 
\begin{equation}
{\cal L}={1\over 2}|\partial_\mu\Phi|^2-{1\over 4}\lambda\left(|\Phi|^2-\eta^2\right)^2\,.
\label{Lagrangian}
\end{equation}
We remove the coupling constant, $\lambda$, and symmetry breaking scale, $\eta$ by scaling energy and length units by $\lambda\eta^4$ and $(\sqrt{\lambda}\eta)^{-1}$, respectively. Effectively, this corresponds to setting $\lambda=\eta=1$ in (\ref{Lagrangian}). The complex field can be written as $\Phi=\phi\exp[i\alpha]$ where the radial field, $\phi=\eta+R$, has mass $m_{\rm R}=\sqrt{2\lambda}\eta$, and the phase $\alpha$ is massless - we will not consider the situation where the axion becomes massive here. There is a static straight string solution with $\phi=\phi(r)$ and $\alpha=\theta$, where $(r,\theta,z)$ are cylindrical polar coordinates. The profile function $\phi(r)$ can be calculated numerically, for example, using the method of Successive Over Relaxation (SOR)~\cite{hadjidimos2000successive}.

\begin{figure}[!t]
    \centering
    \vspace{-2.5\baselineskip}
        \includegraphics[scale = 0.9]{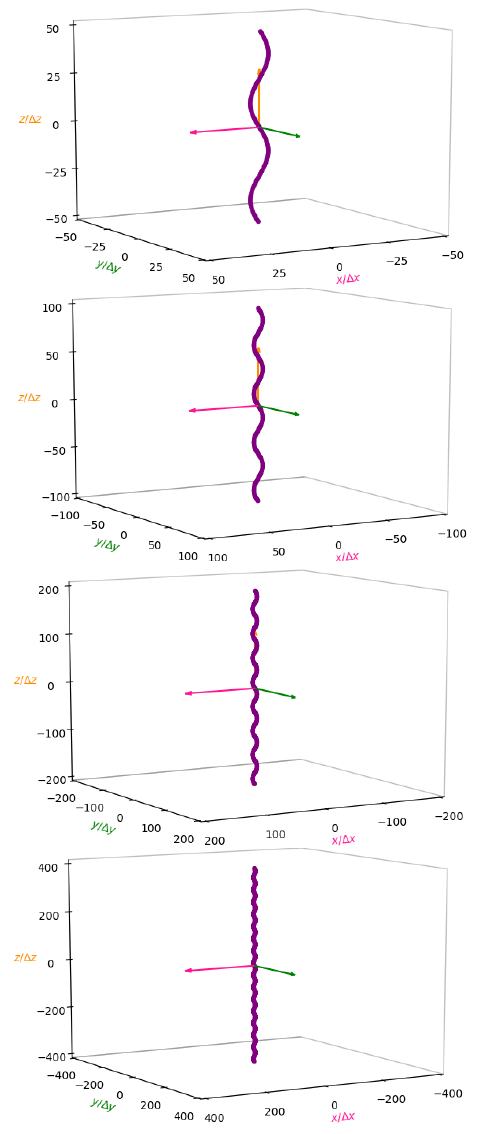}
       
    \caption{Illustrations of the perturbed string configurations for $n_{x}=n_{y}=n_{z}=101,201,401,801$. The positions presented are those which are computed by the string position finding algorithm. The four setups correspond to the same wavelength of $L = 50\Delta x=35$ and $\varepsilon_0=0.5$. Note that the string amplitude is significantly smaller than the box size in all four cases.}
    \label{three_grids}
\end{figure}

\begin{figure}[!h]
    \centering
    \includegraphics[scale = 0.42]{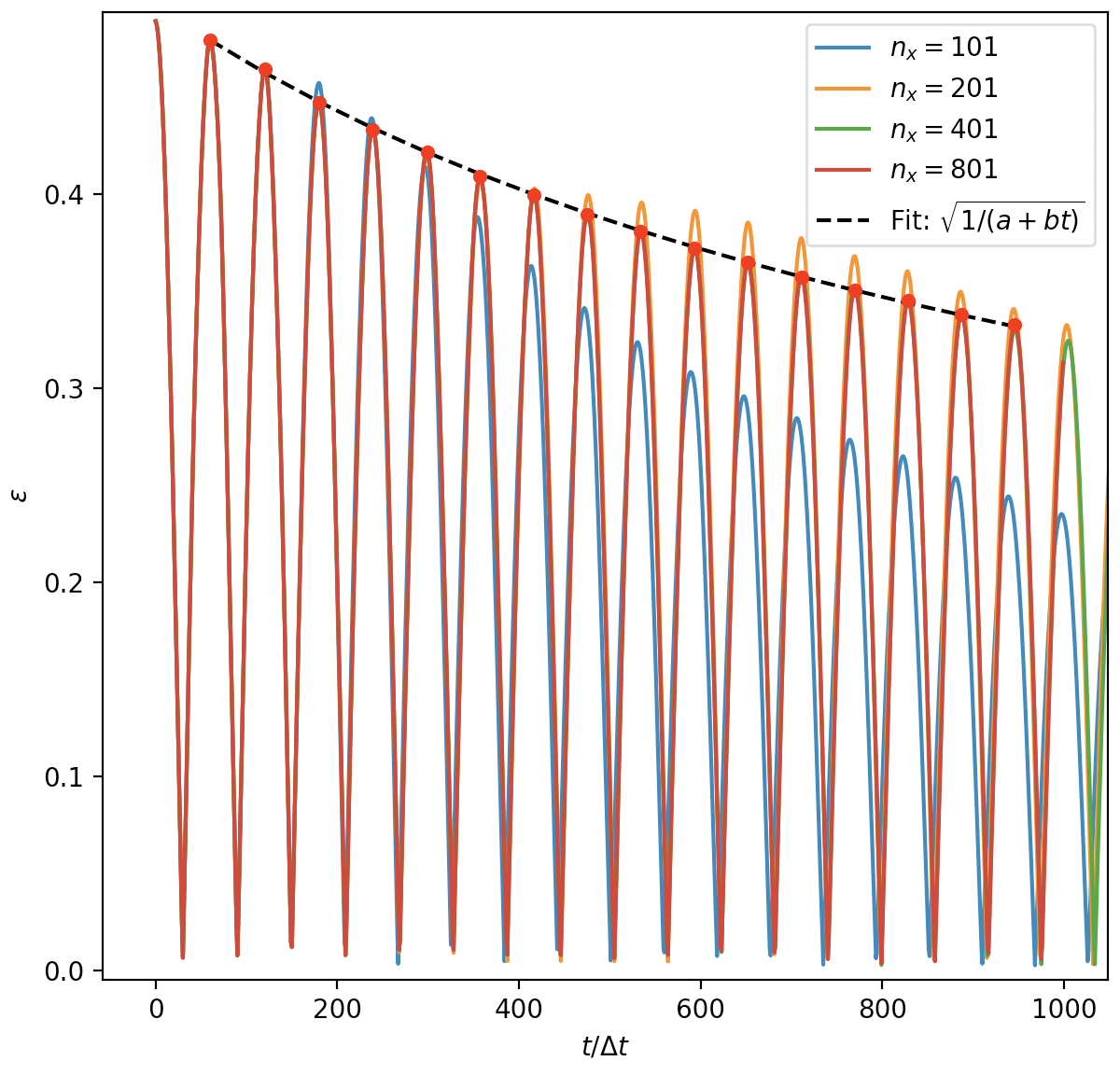}
    \vspace{-1.6\baselineskip}
    \caption{Relative amplitude, $\varepsilon(t)=2\pi A/L$ for the four simulations discussed in the text with $n_x=101, 201, 401$ and 801 for $L=35$ and $\varepsilon_0=0.5$. They are similar for the first few oscillations but deviate after this due the effects of the boundary implying that even with the absorbing boundary conditions there are some reflections of radiation. Included also is a linear fit to $\varepsilon^{-2}$ versus time to the simulation with $n_x=801$.}
    \label{fig:epsilon}
\end{figure}

The equations of motion for the model are discretized on a grid with $n_{x}$, $n_{y}$ and $n_{z}$ points in the $x-$, $y-$ and $z-$directions, respectively. In what follows we will use $n_x=n_y=n_z$ in order to make numerical Fourier transforms easier. We use fourth order differences for the spatial grid - except one point from the boundary - and second order differences for the time evolution. The spatial size is $\Delta x=0.7$ and the time step is $\Delta t=0.3$. In this section we will consider only configurations which are periodic in the $z$-direction and hence we use fixed boundary conditions in that direction. We will use absorbing boundary conditions in the $x-$ and $y-$directions, which solve the equation 
\begin{equation}
   \frac{\partial}{\partial t}\frac{\partial \Phi}{\partial x} - \frac{\partial^2 \Phi}{\partial t^2} + \frac{1}{2}\left(\frac{\partial^2 \Phi}{\partial y^2} + \frac{\partial^2 \Phi}{\partial z^2}\right) = 0\,,
   \label{absorbing_boundary_conditions}
\end{equation}
at a boundary in the $x-$direction and a permuted version ($x\rightarrow y$) in the $y-$direction. This approach heavily suppresses the reflected wave allowing the simulation domain to be effectively infinite in $x-y$ plane with sub-percent reflections~\cite{battye1994global}.

\begin{figure*}
    \centering
\includegraphics[scale=1]{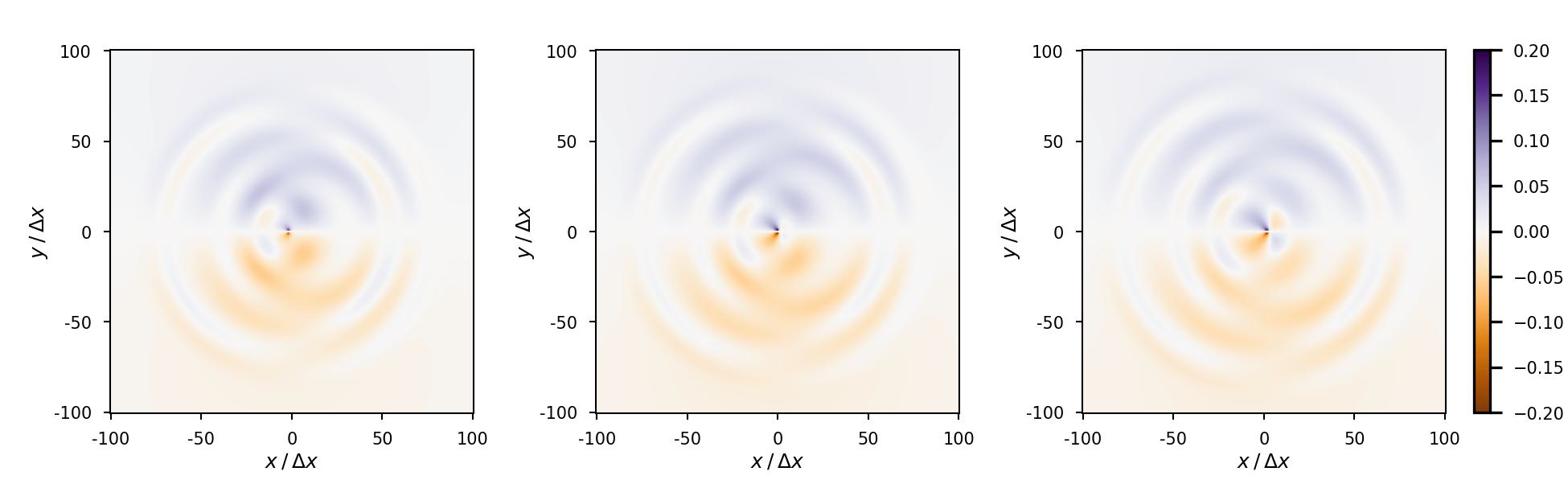}

    \caption{The axion radiation field, $\Delta \alpha$, in the $x-y$ plane for $n_x=801$, $z=412\Delta x$ and $\varepsilon_0=0.5$ for three different time steps (from the left): $t=199\Delta t$, $t=209\Delta t$ and $t=219\Delta t$. We have zoomed in the region in the $x-y$ plane where $-100\le x  /\Delta x,y/\Delta x\le 100$. It is clear that there is a quadrupole pattern near the center of the string. The string was initially displaced in the x-direction. }
    \label{quadrupole}
\end{figure*}

To initialize our simulations for a sinusoidally perturbed string, we use the straight string solution given in the x-y plane by $\Phi=\phi(r)\exp[i\theta]$ where $\phi(r)$ is a profile function calculated numerically, in terms of the radial coordinate, $r$, and angular coordinate, $\theta$. For each z-slice, we offset the origin of this coordinate system in the x-direction according to $A_{0} \sin(2\pi z/L)$ where $A_{0}$ is the initial amplitude and $L$ is the wavelength of the oscillation. The fields are initialized as static; that is the field for two initial time steps is set to be equal.

In Fig.~\ref{three_grids}, we present the sample results of the our algorithm to search for string position for four box sizes with the same wavelength of perturbation, $L=50\Delta x=35$ in dimensionless units.  The one with $n_x=801$ will be used to produce the results presented in this section. Those with $n_x=201$ and $n_x=401$ are reported in appendix \ref{sec:power_calc_appendix} as a comparison, and $n_x=101$ was used in ref.~\cite{battye1994global}. As explained in the appendix it is necessary to use $n_x=801$ in order to clearly see the features in the spectrum of radiation. Note that we also define $\varepsilon=2\pi A/L$ where $A$ is the amplitude of the sinusoidal perturbation and $\varepsilon_0$ will be used to denote the value of $\varepsilon$ at the beginning of the simulation. This was seen to be the relevant quantity in sinusoidal solutions of the Nambu action~\cite{battye1994global,drew2022radiation} and the power per unit length is $\propto \varepsilon^4$. The algorithm works by checking for positions where $\Phi=0$ on the faces of each cube in our grid, of side length $\Delta x$. This is done by using the corners of the face to construct a bilinear interpolation function for $\Phi$, which can be inverted to estimate the point where the string pierces the cube.

In Fig.~\ref{fig:epsilon} we present the evolution of $\varepsilon$ as function of time for the four simulations shown in Fig.~\ref{three_grids}. Initially, the three different size boxes are very similar but they eventually diverge. The case of $n_x=101$ diverges from the other three first around $t=15-20$. Subsequently, the $n_x=201$ and $n_x=401$ cases diverge from the $n_x$ case around $t\approx 40$ and $t\approx 90$. We believe that this is due to imperfections in the absorbing boundary conditions. We take the largest value of $n_x$ to be the one that is most indicative of evolution in the absence of other strings, but it is already clear that subtle effects of the boundaries can induce artifacts on the evolution of the string configurations, and in particular it is not difficult to imagine that other strings and emitted radiation can lead to significant differences in the evolution and, in particular, on the nature of the radiation.

The power emitted by one wavelength of the string is $P=\beta\varepsilon^4$~\cite{battye1994global} and the energy is $E=\mu L\left(1+{1\over 4}\varepsilon^2\right)$. By setting $P=-{\dot E}$ we can deduce that 
\begin{equation}
    {1\over \varepsilon^2}={1\over \varepsilon_0^2}+{4\beta t\over \mu L}=a+bt\,.
\end{equation}
We have fitted the decay of the string amplitude for $n_x=801$ to this and we found $a=4.0$ and $b=0.018$ which implies $\varepsilon_0=0.50$ as it should, and $\beta/\mu\approx 0.16$. The prediction from the KR action is that $\beta\approx\pi^3f_{\rm a}^2/16$ and, hence, we deduce that $\mu\approx 12f_{\rm a}^2$ which is compatible $\mu\approx \pi f_{\rm a}^2\log(L/\delta)\approx 11 f_{\rm a}^2$ if $\delta\approx 1$.

The phase of the field can be written as $\alpha=\alpha_{\rm str}+\Delta\alpha$ where $\alpha_{\rm str}$ is that which corresponds to the string self-field and $\Delta\alpha$ is the axion radiation field which we are most interested in. The key issue that we want to discuss here is separating the two within the simulations. When the oscillating, perturbed string is straight $\alpha_{\rm str}$ will be approximately azimuthal, that is, $\alpha_{\rm str}\approx\theta$. This is the methodology used in refs.\cite{Davis:1989nj,battye1994global}. There will be corrections to this; one of these is due to the fact that the string would be moving and we will return to this later. Numerically, the main error associated with this method is due to the uncertainty of the detected position of the string. This error is most pronounced when the string is aligned with the grid, because in this limit the string is tangent to the faces of the cubes in the grid and, therefore, does not clearly pierce it at a unique position. To avoid this scenario, we perform the subtraction of the string ansatz when the antinodes are displaced by $\approx0.5\,\Delta x$.  

In Fig.~\ref{quadrupole} we present the results of subtracting the static field positioned at the measured centre of the string from the overall phase of the field for three different time steps for single string simulations with $n_x=801$ and $\varepsilon_0=0.5$. These are not all points where the string is straight illustrating that for $\varepsilon<1$ the self field subtraction works very well even when the string is not straight. The emitted radiation exhibits a clear quadrupole pattern in the $x-y$ plane as expected from previous work~\cite{Davis:1989nj,battye1994global}. We see that $-0.2\le\Delta\alpha\le 0.2$ whereas $0\le\alpha< 2\pi$ and the two are clearly very different!

In ref.~\cite{drew2022radiation,drew2023axion, Drew:2022iqz} an alternative way of separating the axion from the self-field of the string is introduced and this is applied to both the massless axion radiation and also the massive radiation associated with the radial field, $R$. This involves calculating the projection of the momentum flux components of the energy-momentum tensor in the radiation direction. If $\Phi=\phi_1+i\phi_2$ then we can define 
\begin{equation}
    {\cal D}_i\alpha={\phi_1\partial_i\phi_2-\phi_2\partial_i\phi_1\over\phi}\,,
    \label{drew_diag}
\end{equation}
and the radiated power in the massless modes can be defined as ${\cal P}_{\alpha}={\hat r}^{i}{\cal D}_i\alpha$ where ${\hat r}^i$ is the radial unit vector in cylindrical coordinates. If $\alpha\approx\theta+\Delta\alpha$, as is assumed by the self-field subtraction, then ${\cal P}_{\alpha}\approx \phi{\hat r}^i\partial_i\Delta\alpha$.

\begin{figure*}

 \vspace{-3.5\baselineskip}

\includegraphics[scale=1]{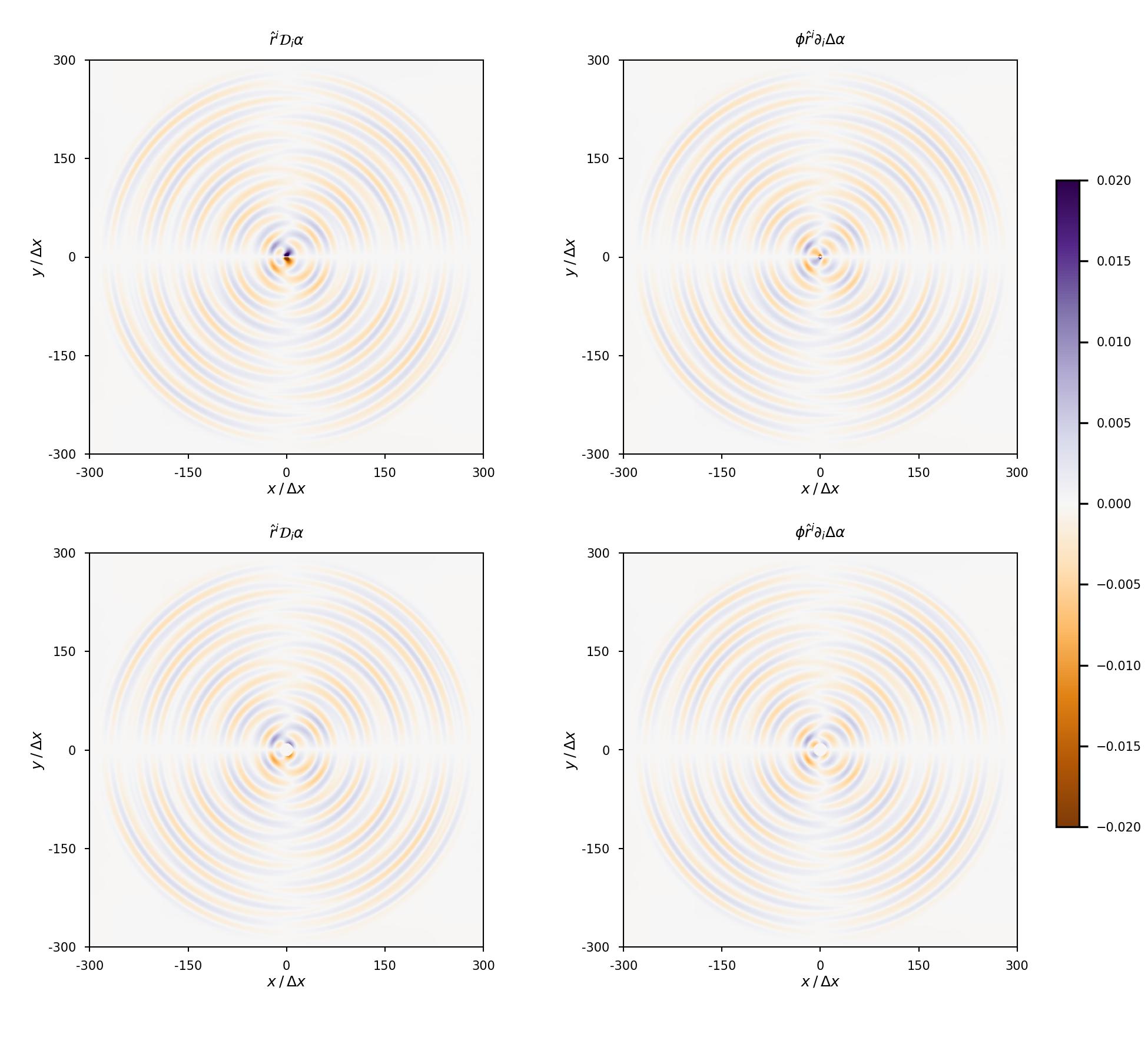}

 \vspace{-1\baselineskip}
\caption{Comparison of $ \hat{r^i} D_i \alpha $ (left) and $ \phi \hat{r^i} \partial_i \Delta \alpha $ (right) for a single string simulation with $n_x=801$, $\varepsilon_0= 0.5$ for the slice where  $z = 412\Delta x $ taken at time $t = 678\,\Delta t$. We have zoomed in the region in the $x-y$ plane where $-300\le x  /\Delta x,y/\Delta x\le 300$. As can be clearly seen the two quantities are almost identical, with small differences very close to the string. After the application of a mask of radius $r = 10\,\Delta x$ (bottom row) the two quantities become identical.}
\label{equating_diagnostic_to_sss}
\end{figure*}

In Fig.~\ref{equating_diagnostic_to_sss} we present a comparison of ${\cal P}_\alpha$ calculated using (\ref{drew_diag}) and $\phi{\hat r}^i\partial_i\Delta\alpha$ calculated using the self field subtraction method described above. We see that the two are very similar except close to the centre of the string. If we apply a cylindrically symmetric mask for $r<r_{\rm crit}\approx 10\Delta x=7$ then the two are visually almost identical and we have confirmed this by performing a spectral analysis in the 2D plane. It is difficult to be sure which of the two methods is most representative of the truth for $r<r_{\rm crit}$. We note that $r_{\rm crit}\sim A=\varepsilon_0L/(2\pi)$ meaning that the region we have removed is more or less compatible with that in which the string is oscillating and where we might not expect either method to work.

The two methods we have discussed above - calculation of $\Delta\alpha$ and ${\cal P}_\alpha$ - have been seen to be very similar and are easily applied in the case of a perturbed straight string. However, in network simulations there are many strings in a small volume with significant radiation, both massless and massive, confusing things. It is difficult - although not necessarily impossible - to decide which string to use to either subtract the static ansatz, or to define the radial unit vector with which to compute ${\cal P}_\alpha$.  

\begin{figure*}
    \centering

\includegraphics[scale=1]{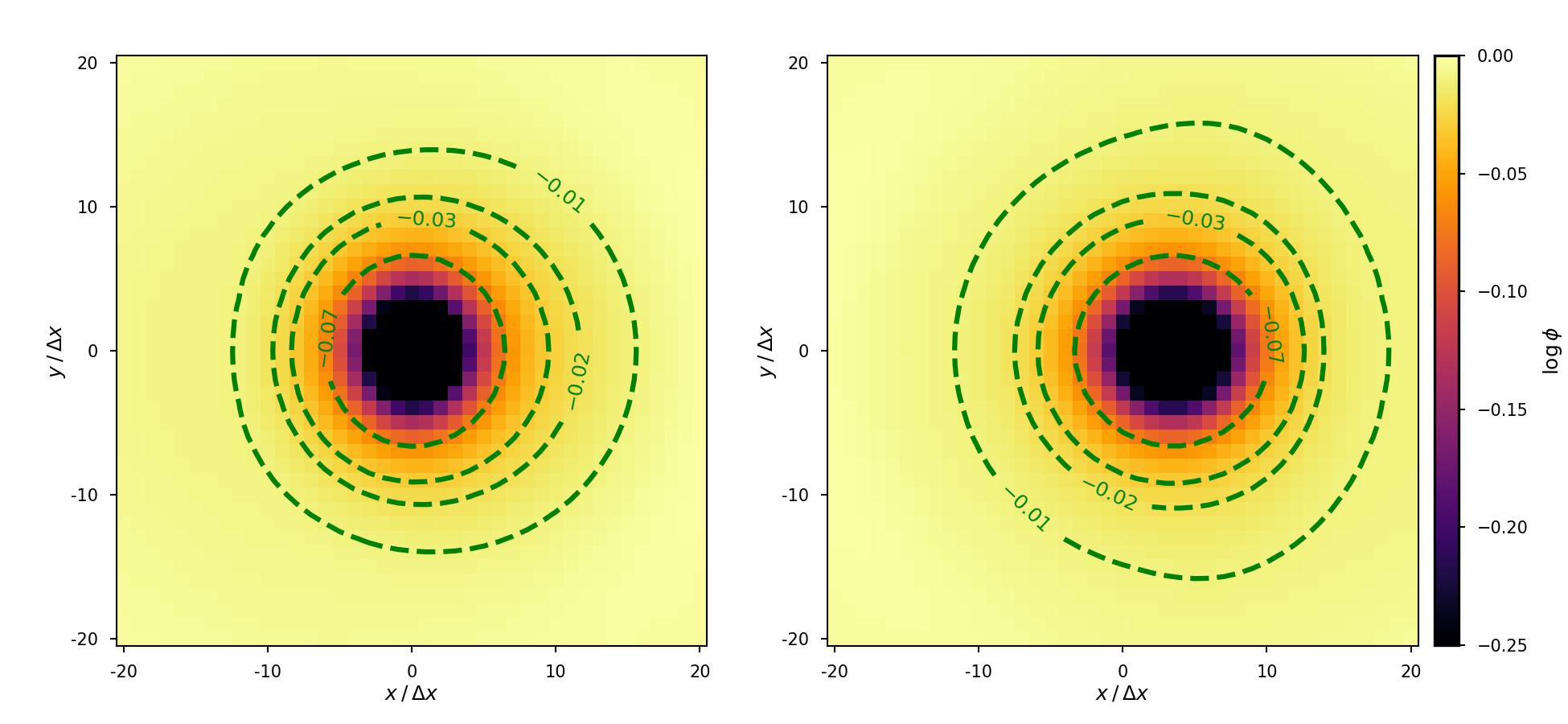}
    \vspace{-1\baselineskip}
    \caption{In the left-hand panel we present the field $\phi$ in the $x-y$ plane at $z=412\,\Delta x$ and $t=209\,\Delta t$ for simulation with $n_{x}=801$ and $\varepsilon_0=0.5$, but we have zoomed into the region $-20\le x  /\Delta x,y/\Delta x\le 20$. The string is at the centre of the grid. In the right-hand panel we present a version of the left panel at a later time $t=239\,\Delta t$, where the string is slightly to the right with respect to the centre of the grid and at its maximal displacement. The effect of the Lorentz contraction of the field near the center of the string is visible when comparing the left panel to the right one and contours are clearly contracted along the axis which is parallel to the direction of the string's motion. Further away from the core of the string the contours are visibly deformed, which shows the non-local character of the field's behavior.}  
    \label{string_cross_section_phis}

\end{figure*}

\begin{figure*}
    \centering

\includegraphics[scale=1]{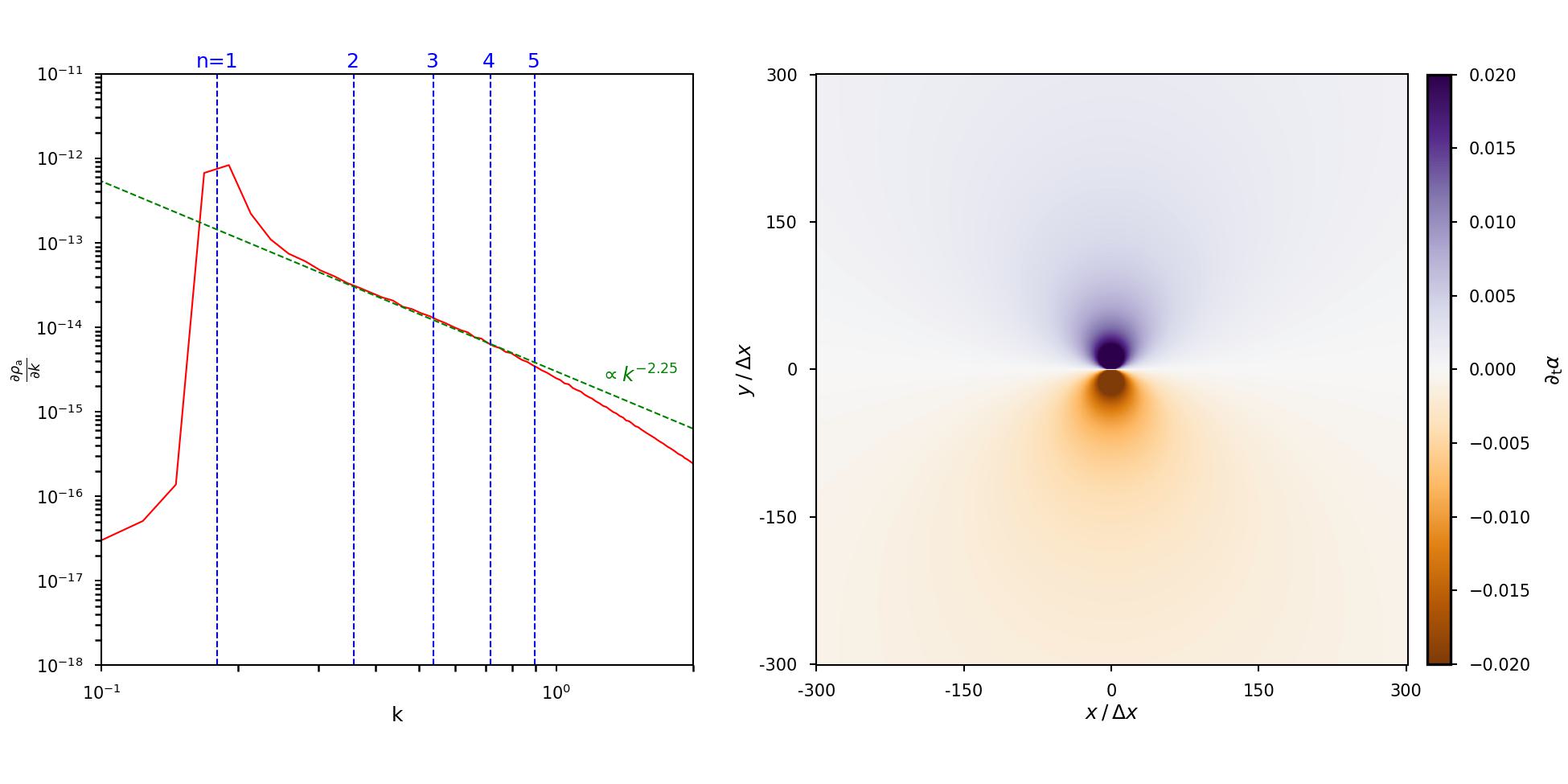}
    \vspace{-1\baselineskip}
    \caption{In the right-hand panel we present the $\phi\partial_{\rm{t}}\theta$, zooming into the region $-300\le x  /\Delta x,y/\Delta x\le 300$ constructed using  (\ref{string_contribution}) for the simulation with $n_{x}=801$ and $\varepsilon_0=0.5$ at $t = 678\,\Delta t$. The field shows a clear dipolar pattern. In the panel on the left we present the spectrum of this quantity together with the positions of the first five harmonics of the string given by $k = 2\pi n/L$. The spectrum peaks around the first harmonic with a power law fall-off with $k^{-2.25}$ as demonstrated by the green dashed line.}  
    \label{string_cross_section_phis_2}
\end{figure*}

\begin{figure*}
    \centering
\includegraphics[scale=1.0]{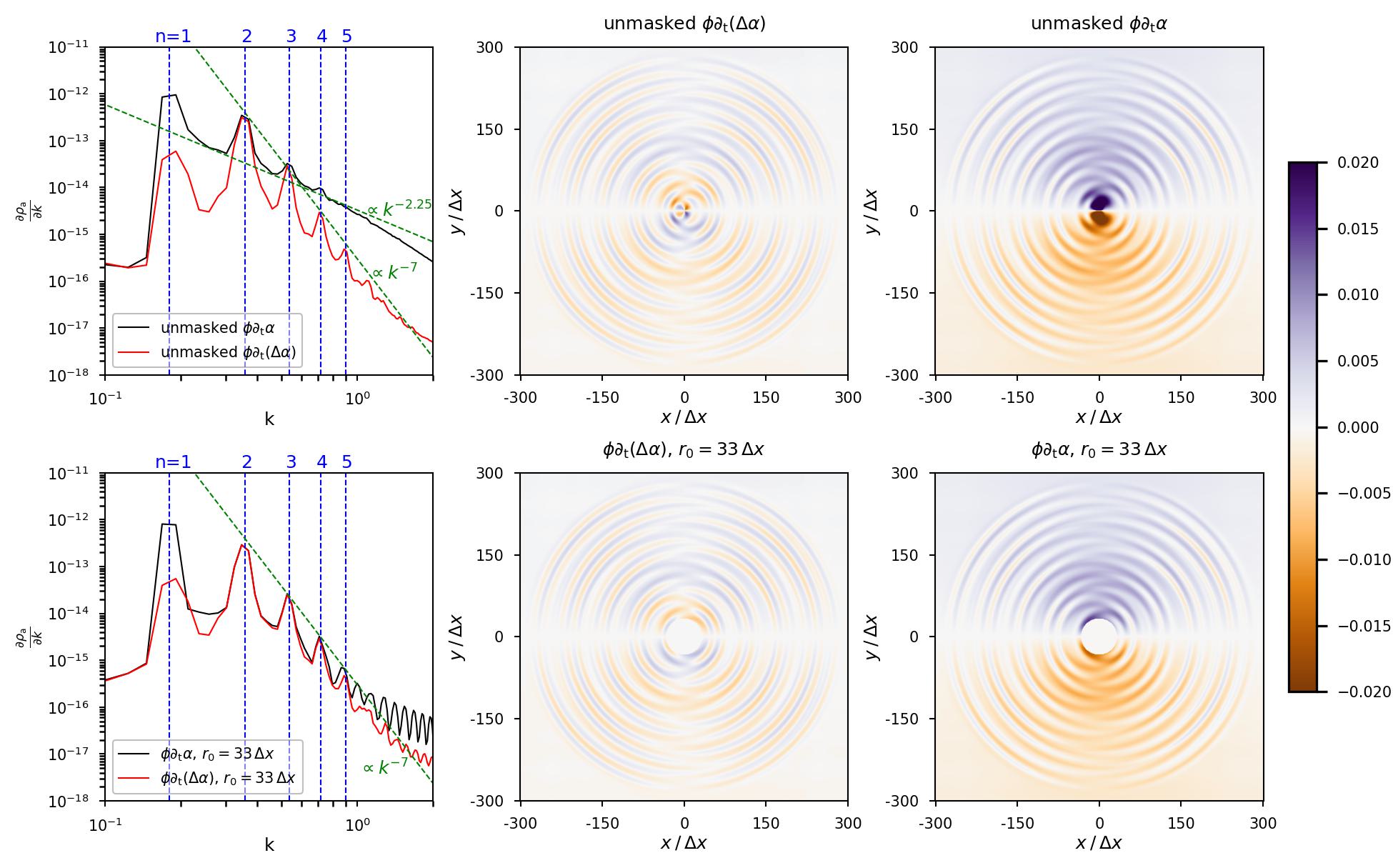}

    \caption{In the top panel of the column on the right we present the axion field $\phi\partial_{t}\alpha$ outputted from the simulation at $t = 678\,\Delta t$ in the $x-y$ plane for $z=412\,\Delta x$. In the bottom panel of the same column we present the same quantity after the mask of radius $r_{0}= 33 \,\Delta x$ was applied. The middle column presents the field after the self-field subtraction i.e. $\phi{\partial_{t}(\Delta\alpha})$. We further present the spectra of these quantities in the column on the left, together with the green dashed lines that show power laws characteristic for each spectrum and the vertical blue dashed lines showing the positions of first five harmonics given by $k = 2\pi n/L$. The spectra in the bottom panel show a clear convergence to a common line between the $n=2$ and $n=5$ harmonics.}
    \label{circular_mask_field}
\end{figure*}

Now we will discuss the basics of the analysis of the axion energy spectrum. If we write $\Phi=\phi e^{i\alpha}$ then energy is given by 
\begin{eqnarray}
    E &=& \int d^3{\bf x} \bigg(\frac{1}{2}\phi^2\Big[(\partial_{t} \alpha) ^2 +\lvert \nabla \alpha \rvert^2\Big] \cr &+&  \frac{1}{2}(\partial_{t} \phi)^2 + \frac{1}{2}\lvert \nabla \phi \rvert ^2 + V(\phi)\bigg) \,.
\end{eqnarray}
Using the decomposition $\alpha = \alpha_{\rm{str}}+\Delta\alpha$, one can write the terms dependent on $\alpha$ as
\begin{equation}
    E_{\alpha} = \int d^{3}\mathbf{x}\bigg( \rho_{\rm{a}}+\rho_{\rm{int}}+\rho_{\rm{str}}\bigg)\,,
\end{equation}
 where 
 \begin{equation}
     \rho_{\rm{a}} = \frac{1}{2}\phi^2\left[(\partial_{t}(\Delta \alpha))^2+\lvert \nabla (\Delta\alpha) \rvert^2\right] \approx\phi^2\left[\partial_{t}(\Delta \alpha)\right]^2\,,
\end{equation}    
corresponds to the energy density of {\em free} axions, while the remaining terms are due to the energy density of the strings and the interactions the axions and the string. The key thing to understand is that contribution from the strings is significant and the energy in axions is not the integral of $\phi^2(\partial_t\alpha)^2$ which is what is typically computed in other works. In what follows we will demonstrate that close to the string this quantity is dominated by the string and not axions; avoiding including this effect is crucial to understanding the spectrum of radiation emitted by the strings given that the relic density is so sensitive to the value of $q$.

The axion energy density can be expressed as  
\begin{eqnarray}
     \rho_{\rm{a}} &=& \frac{1}{L^3}\int  \phi^{2}(\partial_{t}(\Delta\alpha)) ^2 d^{3}\mathbf{x}\cr &=& \frac{1}{(2\pi L)^3}\int  {\widehat {\lvert(\phi\partial_{t}(\Delta \alpha)({\mathbf{k})} \rvert^2}}d^{3}\mathbf{k}\,,
\end{eqnarray}
where to obtain the second equality we use Parseval's theorem and ${\hat f}({\bf k})$ is the Fourier transform of $f({\bf x})$. This expression can be further rewritten as   
\begin{equation}
     \rho_{\rm{a}} = \int dk\frac{k^2}{(2\pi L)^3}\int  \widehat{\lvert\phi\partial_{t}(\Delta \alpha) \rvert}^2 d\Omega_{\rm k} = \int dk\frac{\partial\rho_{a}}{\partial k}\,,
\end{equation}
which allows us to deduce an expression for the spectral density
\begin{equation}
     \frac{\partial\rho_{a}}{\partial k} = \frac{k^2}{(2\pi L)^3}\int  \widehat{\lvert\phi\partial_{t}(\Delta \alpha) \rvert}^2 d\Omega_{\rm k}\,,
     \label{spectrum_def}
\end{equation}
where $k$ is the proper momentum and L is the box size.

To calculate the spectrum of radiated axions, as opposed to the axion field, we need to calculate (\ref{spectrum_def}). We have demonstrated that the field subtraction can be done easily for a perturbed straight string by making the approximation that $\alpha=\theta+\Delta\alpha$ to subtract the component of the self-field as was done in refs.\cite{Davis:1989nj, battye1994global}, and indeed we have shown that this agrees well with the approach of ref.~\cite{drew2022radiation}. Calculating (\ref{spectrum_def}) requires knowledge of the time derivative of $\theta$ that we approximate using 
\begin{equation}
    \partial_{\rm{t}}\theta =\frac{\gamma\, v \,\rm{sin}\theta}{r}\,,
    \label{string_contribution}
\end{equation}
where $r$ is the distance in the $x-y$ plane from the center of the string and $v$ is the velocity of the string in that plane with $\gamma=1/\sqrt{1-v^2}$. This is calculated from the derivative of $\theta^\prime = \tan^{-1}(y^\prime/x^\prime)$ with respect to $t$, where primed coordinates represent the rest frame of the string. The two coordinate systems are related by the Lorentz boost, $t^\prime = \gamma(t-vx)$, $x^\prime = \gamma(x-vt)$, $y^\prime = y$ and $z^\prime=z$. To avoid having to account for retardation effects, we approximate the calculation by setting $\theta^\prime \to \theta$ and $r^\prime \to r$ in the final expression. We present $\phi\partial_t\theta$ computed this way in the right-hand panel of Fig.~\ref{string_cross_section_phis_2} which shows a clear dipolar pattern, and we will soon demonstrate that this is a dominant contribution to $\phi\partial_{\rm{t}}\alpha$ in the region close to the core of the string. In the left-hand panel of Fig.~\ref{string_cross_section_phis_2}, we plot the spectrum of the string's contribution to the axion field, that is, we compute (\ref{spectrum_def}) for $\alpha=\theta$ using (\ref{string_contribution}). In doing this we used velocities obtained by numerical differentiation of string's position for each z-slice. We see that the spectrum peaks around the first-harmonic mode of the string with $k=2\pi/L$, and for higher values of the momentum, it slowly falls off $\propto k^{-2.25}$, effectively being a sizable background for the free-axion contribution.

We now attempt to compare the spectrum calculated using $\phi\partial_t\alpha$ with that for $\phi\partial_t(\Delta\alpha)$ using the approximation
\begin{equation}
    \phi\partial_t(\Delta\alpha)\approx\phi\partial_t\alpha-\phi_{\rm SOR}\partial_t\theta\,,
    \label{time_subtract}
\end{equation}
in Fig.~\ref{circular_mask_field}, where $\phi$ and $\alpha$ are computed from the field in the simulation and $\phi_{\rm SOR}$ is the static field of the string computed using SOR. In the right-hand panel of the top row, we show $\phi\partial_t\alpha$ in the plane $z=412\Delta x$ corresponding to one where the string has maximum. We choose $t=678\Delta t$ which is when the string is actually close to straight, that is, when it is moving at its maximum velocity. The spectrum calculated for the field at this time, using all three dimensions, is presented in the left-hand panel. Visually, the field in the $x-y$ plane is dominated by the centre, where the string is positioned,  and a clear $y>0$ versus $y<0$ asymmetry. The spectrum has a clear peak at $k=2\pi/L$, the first harmonic, and smaller peaks at $k=4\pi/L, 6\pi/L, 8\pi/L$ corresponding to the $n=2,3,4$ modes, plus a fall-of envelope $\propto k^{-2.25}$. It is very clear that the part of the spectrum in the vicinity of the $n=1$ mode is almost identical to the spectrum of $\phi\partial_t\theta$ presented in Fig.~\ref{string_cross_section_phis_2}, and the high frequency tail falls-off almost identically as that of the spectrum of $\phi\partial_{\rm{t}}\theta$. 
\begin{figure*}
    \centering
\includegraphics[scale=1]{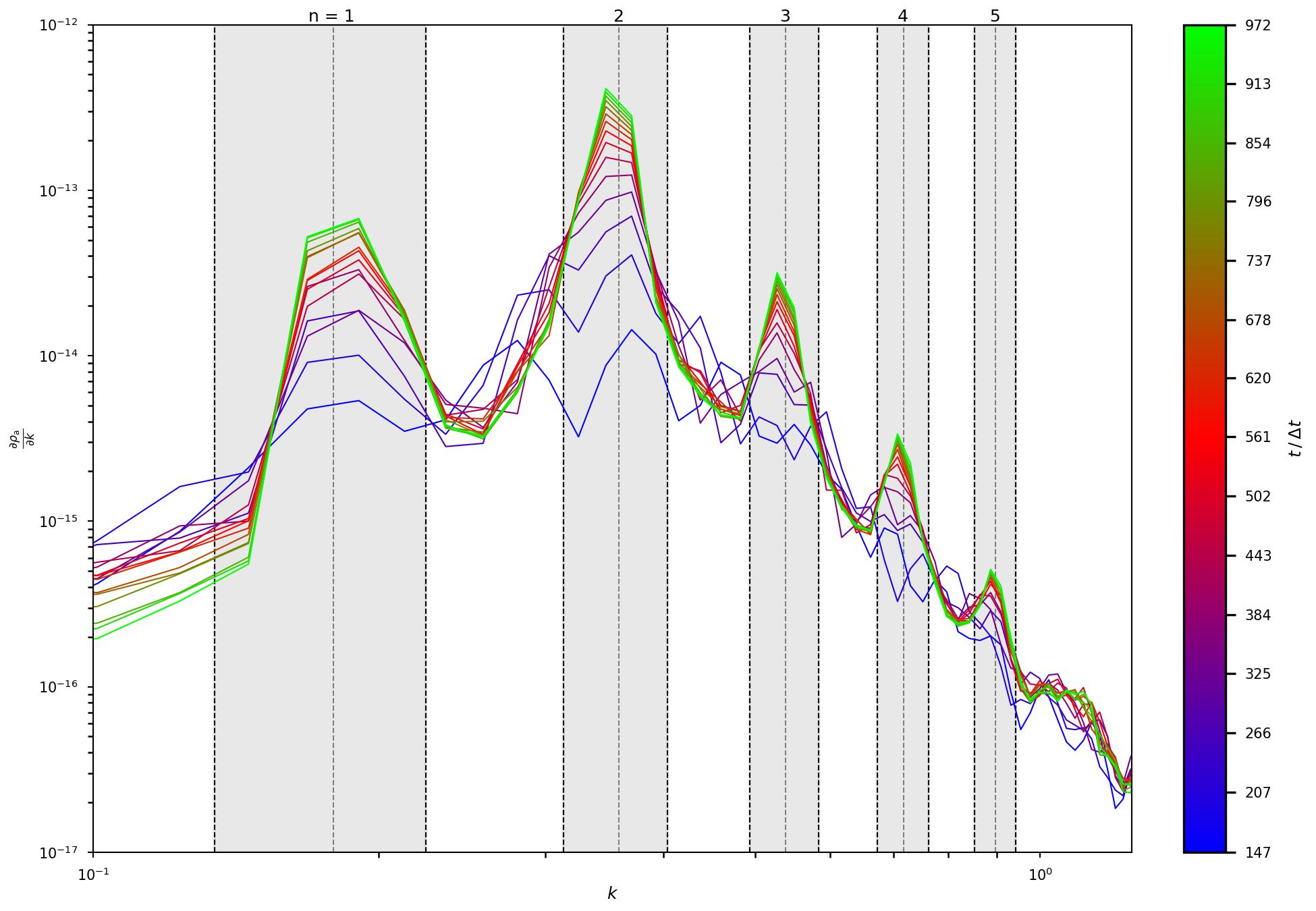}

    \caption{The evolution of the axion energy spectrum (\ref{spectrum_def}) for a simulation with $n_{x} = 801$ and $L = 50\,\Delta x$ after applying a circular mask of radius $r_{0} = 33\,\Delta x$. The spectra are shown from each time step when the string is approximately straight, from the second oscillation until and including the first such time step after the radiation hits the boundary (that happens at $t =n_x\Delta x/2$). The grey dashed lines show the first five harmonics of the string enumerated with $n$ and with momenta given by $k =  2\pi n/L$. Black dashed lines show bands of width $0.09$ centered around the harmonics. The formation and steady growth of the peaks centered at the harmonic and contained within the bands is visible, while the spectrum outside the bands doesn't appear to be growing in time.}
    \label{spectrum_evolution}
\end{figure*}
In the middle panel, we present the equivalent of the right-hand panel, but using (\ref{time_subtract}) and the spectrum of this is also presented in the left-hand panel. It is clear that using the subtraction suppresses the effect of the strings motion and sharpens the peaks. In the power spectrum the $n=1$ mode is suppressed by around an order of magnitude and the $n=2,3,4$ modes stick up more clearly above the background from the residual due to the imperfections in the subtraction. The heights of the three visible harmonics appear to reduce with a high-power, estimated visually to be approximately $\propto k^{-7}$. The residual imperfections are due to the Lorentz contraction of the field profile $\phi(r)$ already pointed out in Fig.~\ref{string_cross_section_phis}. These effects are difficult to mitigate due to retardation effects. We note that $\varepsilon_0=0.5$ is a relatively extreme example and the subtraction will work much better for lower values of $\varepsilon_0$, although the radiation would be much weaker.

One might be concerned that the subtraction is not perfect near the centre of the string. In the bottom row of Fig.~\ref{circular_mask_field} we show plots that are very similar to the top row, but here we have extracted a cylinder of radius $r_0=33\Delta x\gg A$ along the $z$-axis. In order to correct for the effect of this ``masking'' on the spectrum we have used appendix~\ref{sec:power_calc_appendix}. The radius is chosen so that the spectrum of $\phi\partial_t\alpha$ and $\phi\partial_t(\Delta\alpha)$ agree for $4\pi/L\lesssim k\lesssim 10\pi/L$ and the peaks for $n=2,3,4$ and $5$ appear to reduce following $\propto k^{-7}$, same as the peaks in the spectrum of $\phi\partial_{\rm{t}}(\Delta\alpha)$. There is still a significant residual of the $n=1$ mode in the spectrum of $\phi\partial_t\alpha$ and this can be seen visually in the bottom right-hand figure. 
\begin{figure*}
    \centering
\includegraphics[scale=1]{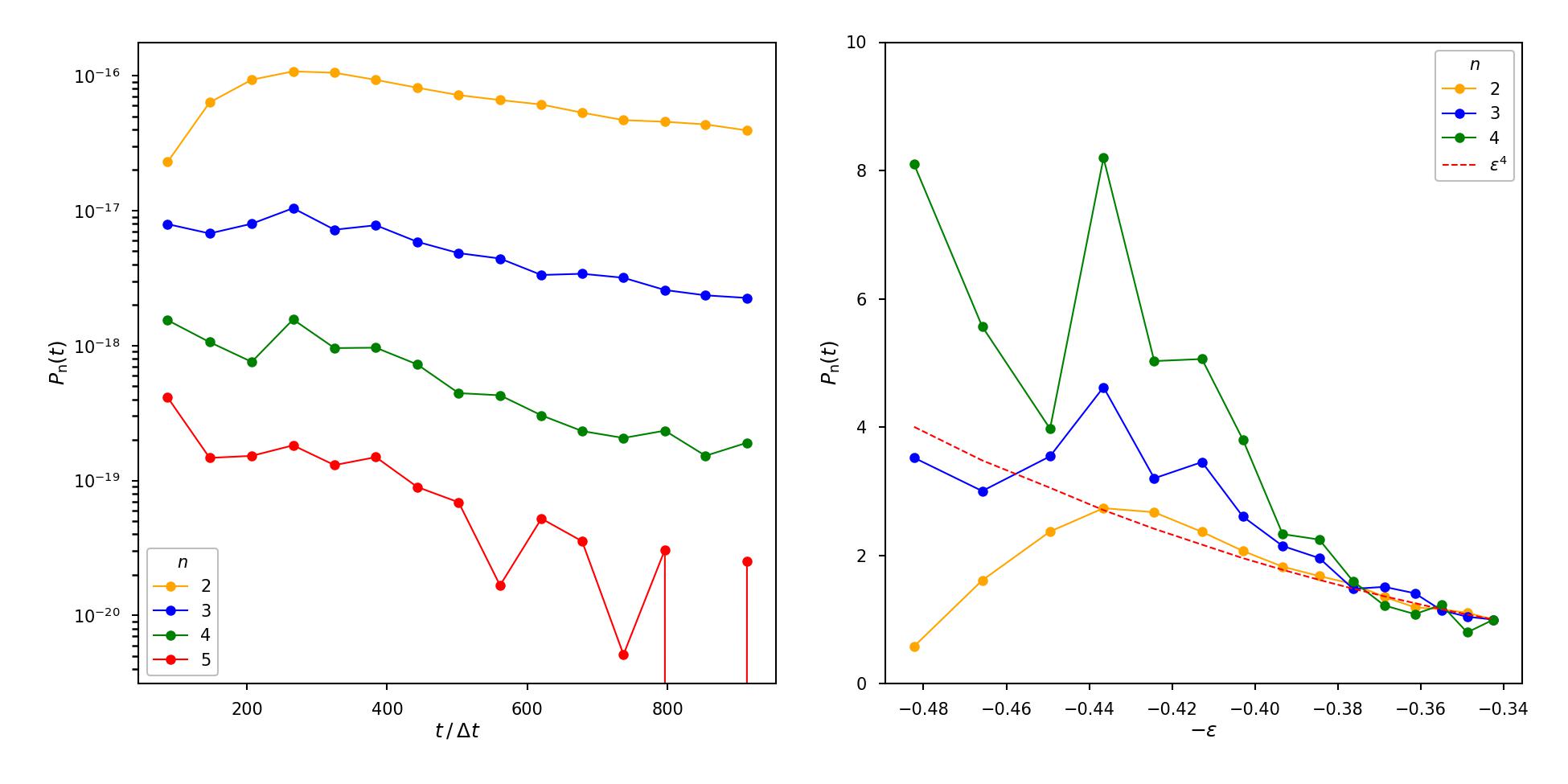}

    \caption{In the left-hand panel, we present the power emitted in each of the peaks shown in Fig. \ref{spectrum_evolution} for the simulation with $n_{x} = 801$ and $L = 50\,\Delta x$. The power is defined as a time derivative of the energy, which in turn is the integral of the energy spectrum within the band of width 0.09 centered around the peak. The power emitted in each of the peaks differ by an order of magnitude and as can be seen in Fig. \ref{spectrum_evolution} the majority of the energy is emitted in the $n=2$ mode, which grows the fastest. In the right-hand panel, we present the comparison of $P\propto\varepsilon^4$ to the power emitted in each mode (now normalizing to the last time step). To convert time to epsilon we linearly extrapolate between the peaks of the line in Fig. \ref{fig:epsilon}.}    \label{q_calc}
\end{figure*}
A conclusion of this work is that the spectrum calculated from $\phi\partial_t\alpha$, as is often the case in numerical field theory simulations of string networks, is dominated by the motion of the string and that even a substantial masking does not remove this, although it sharpens the higher $n$ modes. When a subtraction and masking are implemented we see a sharp fall-off of the spectrum at least from this specific configuration. The dominance of the $n=2$ mode is predicted analytically~\cite{battye1994global} and it is equivalent to the visual appearance of a quadrupole in Fig.~\ref{quadrupole}. We note that a number of authors~\cite{saikawa2024spectrum, Gorghetto:2018myk}, have performed a masking of the field in network simulations and have claimed little impact on their measurements of the spectrum. However, the radii they have used are very much lower than we found necessary above. Moreover, it is clear that the impact of the n=1 will dominate the spectra for a network comprising of long strings/loops with multiple wavelengths/lengths.

We now attempt to estimate the power of axion radiation from the perturbed periodic string configuration. In Fig.~\ref{spectrum_evolution} we present the time evolution of ${\partial\rho_{\rm a}\over\partial k}$ - calculated using the field subtraction and the masking with $r_0=33\Delta x$ - for the same initial configuration as in Fig.~\ref{circular_mask_field} and we have measured the power spectrum in bins centred around $n=1-5$ modes - denoted $E_{\rm{n}}(t)$. We then numerically differentiate $E_{\rm{n}}(t)$ with respect to time, using a first order central derivative to extract the power emitted in each bin, that we denote by $P_{\rm{n}}(t)$ and present in the left-hand panel of Fig.~\ref{q_calc}. It is clear that the $n=2$ mode dominates the spectrum - with the $n=1$ mode (not shown) being strongly present throughout due to the imperfection of the subtraction - and that the modes with higher values of $n$ emerge with time. We note that since the envelope of $\varepsilon$ shown in Fig.~\ref{fig:epsilon} is a monotonically decreasing function of time, we can convert the axis from time to $\varepsilon$, where the values for the envelope are obtained by linear extrapolation between the neighboring peaks in Fig.~\ref{fig:epsilon}. When we change variables, the $P_{\rm{n}}$ (normalized to the last point) is presented in right-hand panel of Fig.~\ref{q_calc}, together with the line showing $P= \sum_{n} P_{\rm{n}}\propto\varepsilon^4$ relation. We see, that as $\varepsilon\rightarrow0$, the power for each of the bins $n=2,3,4$ tends to $\propto\varepsilon^4$. Hence, we verified the prediction that in a limit $\varepsilon\rightarrow0$, the power emitted by the string is $\propto\varepsilon^4$.
\begin{figure*}
    \centering
\includegraphics[scale=1]{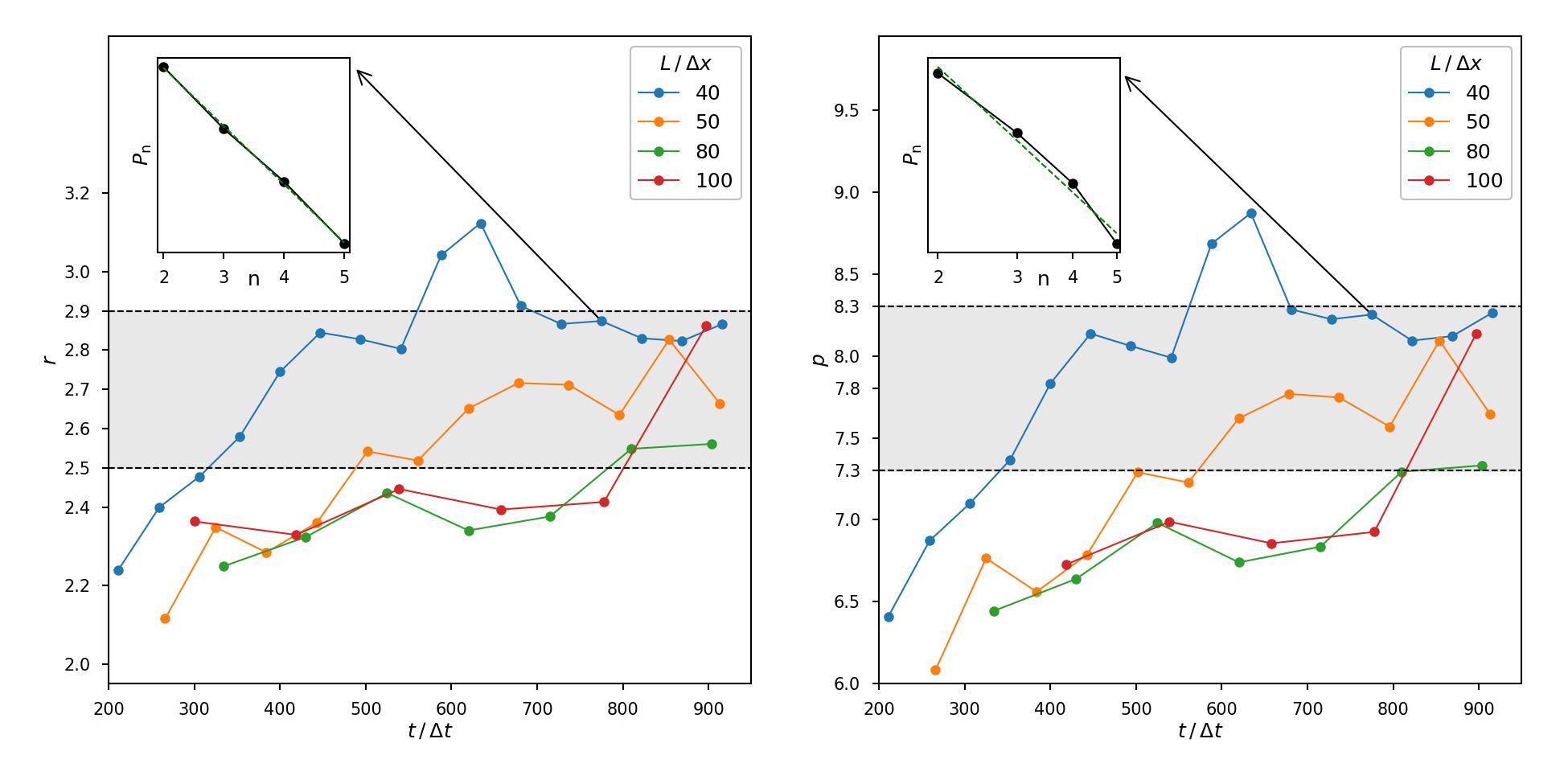}

    \caption{In the left-hand panel we present the evolution of the parameter $r$ for the simulations with $n_{x} = 801$, $\varepsilon = 0.5$ and $L = 40\,\Delta x$, $L = 50\,\Delta x$, $L = 80\,\Delta x$, $L = 100\,\Delta x$ estimated using  $P_{\rm{n}}$ for $n = 2,3,4$ shown in Fig \ref{q_calc}. The value of $r$ grows with time and the rate of growth clearly depends on $L$ --- faster growth for simulations with smaller $L$. We also note that the parameter $r$ for the strings with $L = 40\,\Delta x, 50\,\Delta x$ appears to reach a stable state with a value $\approx2.7-2.8$ and for all simulations it reaches the band $2.5-2.9$, by the end of the simulation. In the right-hand panel we present the result of fitting a power law $P\propto n^{-p}$ instead. The behavior of this parameter follows very closely that of $r$ shown on the left-hand side and for all simulations $p$ reaches the band $7.3-8.3$, by the end of the simulation. The inset plots show the fit at an example point (for $L = 40\,\Delta x$ and $t = 775\,\Delta t$ and using $n = 2,3,4,5$). The exponential appears to fit the data better than the power law.}    \label{r_t}

\end{figure*}

We proceed by fitting the relation $P_{\rm{n}}\propto \exp[-{rn}]$ separately at each time step for simulations with $L = 40\,\Delta x,\, 50\,\Delta x,\, 80\,\Delta x,\, 100\,\Delta x$, all chosen so that they correspond to a full number of oscillations in a box with $n_{x}=801$. In this way, the only assumption we need to make is the dependence of $P_{\rm{n}}$ on $n$, while we allow parameter $r$ to be a function of time. The results of this fit are presented in Fig.\ref{r_t}. The value of $r$ for the simulations with smaller $L$ appears to reach a stable value at late times, while the simulations with $L = 80\,\Delta x,\, 100\,\Delta x$ didn't reach this stage, probably due to the string completing fewer oscillations by the end of the simulation. We expect that the values of $r$ for these simulations would eventually become constant, but that would require a larger simulation. To accommodate the uncertainty introduced by the simulations with larger values of $L$, we estimate $r\approx 2.5-2.9$. However, if we were to assume that $P_n\propto n^{-p}$, as is usually assumed for the network simulations, then the best fit is obtained for $p\approx 7-8\gg 1$, but the data appears to fit an exponential much better, as illustrated by the inset plots in Fig. \ref{r_t}.

The value of the spectral index $p$, extracted here, is much larger than $p\leq1$, which is typically obtained from the network simulations ~\cite{Gorghetto:2020qws,saikawa2024spectrum,Benabou:2024msj,buschmann2022dark, Kim:2024wku,Correia:2025nns}. As we showed before, the self-field contributes mainly to the $n=1$ peak of the spectrum $\phi\partial_{\rm{t}}\alpha$, and dominates the rest of the spectrum. One can imagine that, for a network of strings with a distribution of oscillation wavelengths and string velocities, the self-field contribution is not constrained to a single mode but affects a broader range of momenta. Based on the discussion presented in this section, we find it likely that the spectra extracted from network simulations, and the values of the spectral index inferred from them, are dominated by the self-field.
 
\section{Estimates of the relic axion density from strings}
\label{sec:estimates}

\begin{figure*}[!th]
    \centering
    \includegraphics[scale = 0.72]{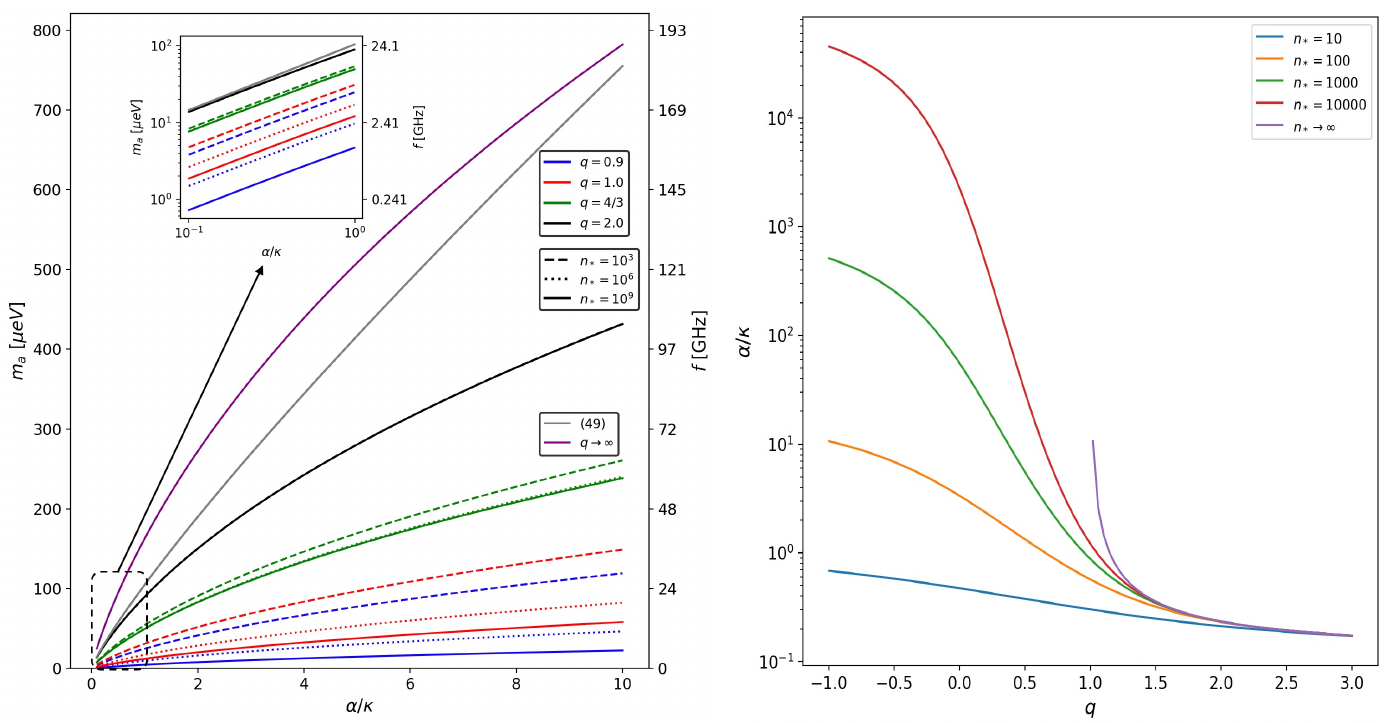}
    \caption{In the left-hand figure we present predictions for the axion mass, $m_{\rm a}$, and detection frequency for ${\cal F}_\ell=1$ as a function of $\alpha/\kappa$ for $\Omega_{\rm a}h^2=0.12$, ${\hat\Gamma}_{\rm a}=46$, $
    \zeta=13$ and $\langle v^2\rangle=0.35$. We have included predictions for $q=0.9,1.0,4/3,2$ and $n_*=10^3, 10^6, 10^9$ and the approximation (\ref{app_mass}) for $q=2$ is shown in gray Note that the approximation converges for $\alpha/\kappa\ll 1$ but is an upper bound for $\alpha/\kappa\gg1$. The main figure uses a linear scale for $\alpha/\kappa$ but there is also an inset for the range $\alpha/\kappa=0.1-1.0$ plotted log-log to illustrate lower values more accurately. Note that for lower values of $q$ there is significant dependence on $n_*$ whereas for $q=2$ there is no visible effect of varying $n_*$. The case of $q=2$ is the closest to the previous analyses in refs.~\cite{Battye:1994au,battye1997recent} albeit for a lower value of $\Omega_{\rm a}h^2$ and with improved calculation for larger $\alpha/\kappa$. In the right-hand figure, we have calculated the value of $\alpha/\kappa$ for which ${\cal R}_{\rm M}=1$, that is where the contribution to the relic density from string loops only is the same as that from the misalignment mechanism when $\langle\theta_i^2\rangle=\pi^2/3$, as a function of $q$ for a range of values of $n_*$. When $n_*=\infty$, one is forced to only consider the region where $q>1$. Just to be clear ${\cal R}_{\rm M}<1$ below the lines and ${\cal R}_{\rm M}>1$ above.}
    \label{fig:fl1}
\end{figure*}

\begin{figure*}[!th]
    \centering
    \includegraphics[scale=0.55]{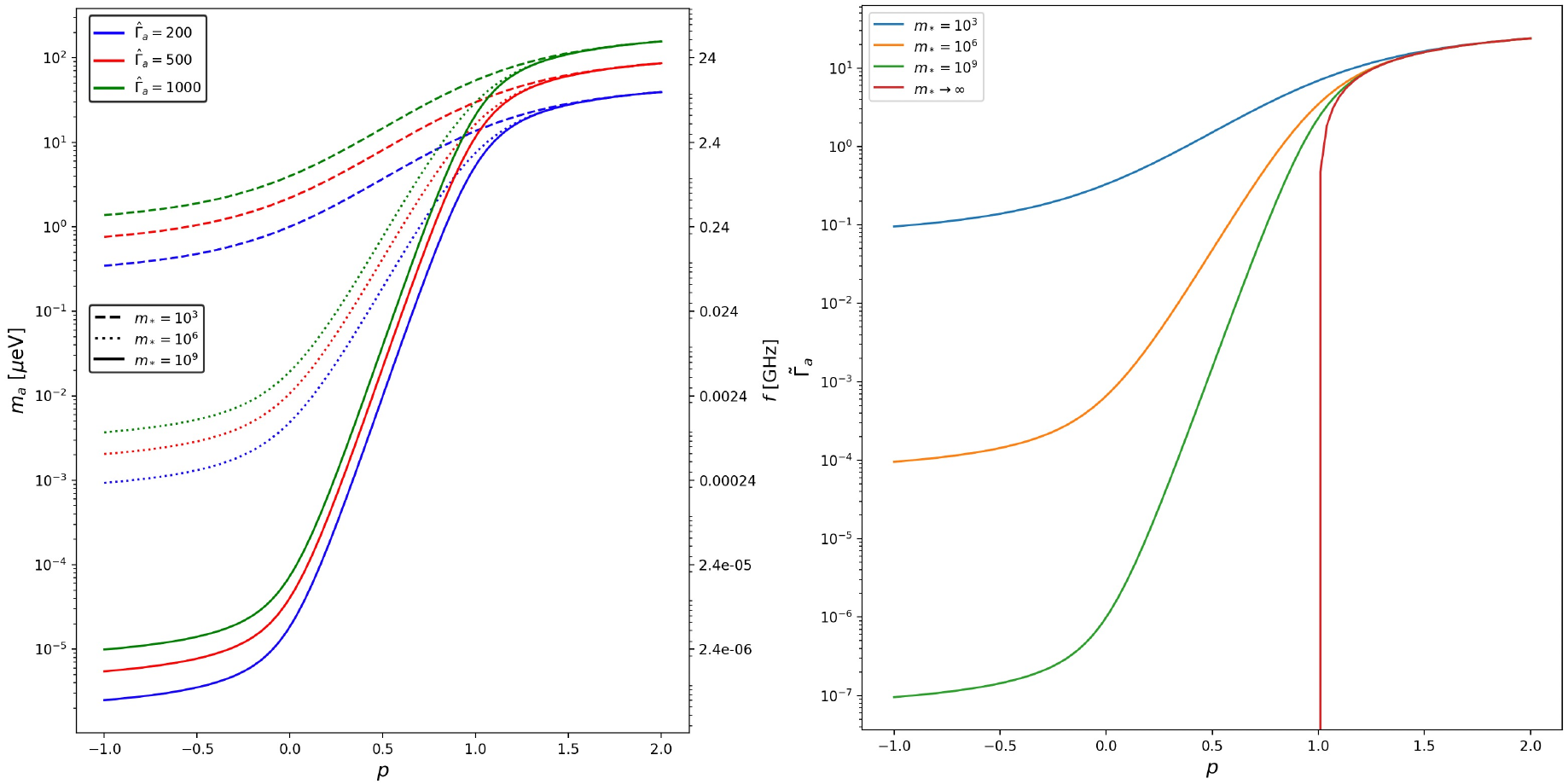}
    \caption{In the left-hand panel we present the predictions for the axion mass, $m_{\rm a}$ and detection frequency for ${\cal F}_\ell=0$ as a function of $p$ for $\Omega_{\rm a}h^2=0.12$, $\zeta=1$ and $\langle v^2\rangle=0.35$. We have included predictions for ${\tilde {\Gamma}_{\rm a}}=200, 500,1000$ and $m_*=10^3, 10^6, 10^9$. In the right-hand figure, we have calculated the value of ${\tilde{\Gamma}_{\rm a}}$ for which ${\cal R}_{M}=1$, that is where the contribution to the relic density from string loops only is the same as that from the misalignment mechanism when $\langle\theta_i^2\rangle=\pi^2/3$, as a function of $p$ for a range of values of $m_*$. When $m_*=\infty$, one is forced to only consider the region where $p>1$.}
    \label{fig:fl0}
\end{figure*}

\begin{figure}
    \centering
    \includegraphics[scale = 0.32]{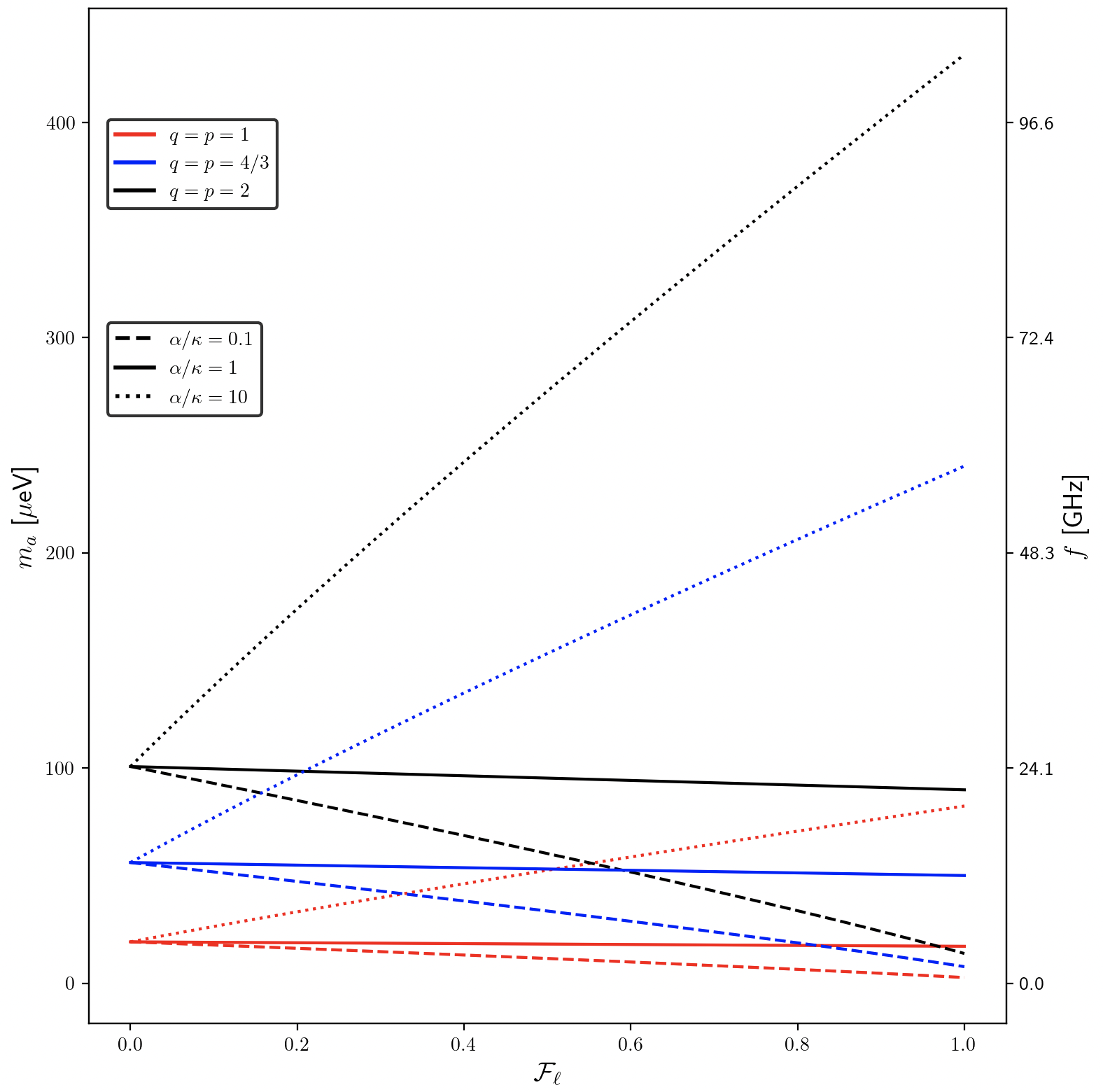}
    \caption{Predictions for the axion mass, $m_{\rm a}$ and frequency as a function of ${\cal F}_\ell$ ranging from $0$ to $1$ for $\Omega_{\rm a}h^2=0.12$, ${\hat \Gamma}_{\rm a}=46$, ${\tilde\Gamma}_{\rm a}=500$, $\zeta=13$ and $\langle v^2\rangle=0.35$. This is presented for $q = p = 1$ (red), $4/3$ (blue) and $2$ (black) for a selection of $\alpha / \kappa$ values of $0.1, 1, 10$ and $n_*=m_*=10^6$. }
    \label{fig:varyfl}
\end{figure}

In the previous section, we have concluded that there are significant uncertainties in the modeling of global string dynamics in numerical field theory simulations. Although we do not necessarily claim that the very best simulations are wrong, we feel that there is sufficient uncertainty that, when trying to predict the axion mass in topological defect-based scenarios, it  is best to resort back to the analytic estimate based on (\ref{overall}) and attempt to assess the uncertainties in certain specific scenarios.

The temperature-dependent axion mass has been estimated to be given by~\cite{DiLuzio:2020wdo} 
\begin{equation}
    m_{\rm a}(T)=\beta_1m_{\rm a}\left({{\bar T}\over T}\right)^{\beta_2}\,,
    \label{eq: DIGA mass}
\end{equation}
around the QCD phase transition where $\beta_1\approx 0.026$, $\beta_2\approx 4$ and ${\bar T}\approx 160\,{\rm MeV}$. By setting $3H(T_{\rm AD})=m_{\rm a}(T_{\rm AD})$, we can deduce that
\begin{eqnarray}
    {T_{\rm AD}\over{\bar T}}\approx 7.0\left({m_{\rm a}\over 360\,\mu{\rm eV}}\right)^{\textstyle{1\over 6}}\left({{\cal N}\over 60}\right)^{-\textstyle{1\over 12}}\,.
    \label{tad}
\end{eqnarray}

If we assume ${\cal N}_{S}\approx{\cal N}$, which is a good approximation near the QCD phase transition, then we can invert the expression for the relic density to give
\begin{equation}
    m_{\rm a}=3.0\,\mu{\rm eV}\left({{\cal N}\over 60}\right)^{-\textstyle{1\over 2}}\left({{\bar T}\over T_{\rm AD}}\right)\left({0.12\over\Omega_{\rm a}h^2}\right){\bar m}_{\rm a}\,,
 \end{equation}
 where ${\bar m}_{\rm a}={\bar m}_{\rm a}(\zeta,\langle v^2\rangle,{\cal F}_\ell,{\hat \Gamma}_{\rm a},{\tilde\Gamma}_{\rm a},\alpha/\kappa,q,n_*,p,m_*)=\zeta(1-\langle v^2\rangle)M$ and $M$ is defined in (\ref{overall}). If we now use (\ref{tad}) then we obtain 
\begin{equation}
    m_{\rm a}=1.1\,\mu{\rm eV}\left({{\cal N}\over 60}\right)^{-\textstyle{5\over 14}} \left({0.12\over\Omega_{\rm a}h^2}\right)^{\textstyle{6\over 7}}{\bar m}_{\rm a}^{\textstyle{6\over 7}}\,.
 \end{equation}
In what follows we will assume ${\cal N}=60$ and $\Omega_{\rm a}h^2=0.12$ before commenting on the impact of possible uncertainties coming from these factors and indeed more generally from the calculation of $T_{\rm AD}$.

Let us first consider the case where ${\cal F}_\ell=1$, that is, just axions produced by loops, which is the case previously studied in refs.~\cite{Battye:1994au,battye1997recent}, then there is no dependence on ${\hat\Gamma}_{\rm a}$, $p$ and $m_*$. If we use $\zeta=13$, $\langle v^2\rangle=0.35$, ${\hat\Gamma}_{\rm a}=46$ then we obtain a predictions for $m_{\rm a}$ as a function of $\alpha/\kappa$ and various values of $q$ and $n_*$ as presented in the left-hand panel of Fig.~\ref{fig:fl1}, as well as the observation frequency deduced using $f_{\rm obs}=m_{\rm a}/(2\pi)$. For $q\gg 1$, $n_*\gg 1$ and $\alpha/\kappa\ll 1$ we have that
\begin{equation}
    {m_{\rm a}\over 190\,{\rm \mu eV}}\approx {f_{\rm obs}\over 44\,{\rm GHz}}\approx \left[{\alpha\over\kappa}\left(1-{1\over q}\right)\right]^{6/7}\,.
    \label{app_mass}
\end{equation}
which is compatible with the inset figure of the left-hand panel in Fig.~\ref{fig:fl1}. Note that the previous analyses of this case used larger values of $\Omega_{\rm a}\approx 1$ since that was the prevailing consensus at the time prior to the discovery of cosmic acceleration. This would have led to lower predicted values for $m_{\rm a}$ and $f_{\rm obs}$ if all parameters describing the decay of the loops were the same. The case of $q=2$ is the closest of the values plotted to what was previously assumed and we see that in that case that there is very little dependence on $n_*$, as one would expect. In addition, we have that, for fixed $\alpha/\kappa$ the predicted values of $m_{\rm a}$ and $f_{\rm obs}$ are larger than for values of $q\sim 1$, and indeed predictions for those values of $q$ are very sensitive to the value of $n_*$. Remembering our earlier discussions concerning scenarios A and B, it is worth noting that scenario A is closest to $q\approx 2$ and scenario B to $q\approx 1$, and hence scenario A typically predicts higher values of $m_{\rm a}$,  all other things being equal, but that scenario B has a significant dependence on $n_*$, whereas for scenario A it is much weaker.

In addition, we present the values of $\alpha/\kappa$ for which ${\rm R}_{\rm M}=1$ for $\langle \theta_i^2\rangle=\pi^2/3$ in the right-hand panel of Fig.~\ref{fig:fl1} using the same values for $\zeta$, $\langle v^2\rangle$ and ${\hat\Gamma}_{\rm a}$. What we see is that the value of $\alpha/\kappa$ where the transition from ${\cal R}_{M}<1$ to ${\cal R}_M>1$ takes place is between 0.1 and 1 for $q\gtrsim 1.5$, but that this starts to increase substantially as $q$ approaches one. This means the loop contribution is typically larger than any possible misalignment contribution unless $\alpha/\kappa$ is small in the regime we have argued for in this paper where $q>1$.

Now consider the case of ${\cal F}_\ell=0$ with $\zeta=1$, $\langle v^2\rangle\approx 0.35$ which is closer to what is seen in the string network simulations. In the left-hand panel of Fig.~\ref{fig:fl0}, we present $m_{\rm a}$ and $f_{\rm obs}$ as a function of $p$ for $m_*=10^3, 10^6, 10^9$ and ${\tilde\Gamma}_{\rm a}=200,500,1000$. Using ${\tilde\Gamma}_{\rm a}=500$ and assuming $p\gg 1$ and $m_*\gg 1$, we find that 
\begin{equation}
    {m_{\rm a}\over 160\,{\rm \mu eV}}\approx {f_{\rm obs}\over38\,{\rm GHz}}\approx \left(1-{1\over p}\right)^{6/7}\,.
\end{equation}
The predictions for $m_{\rm a}$ are comparable to those for ${\cal F}_{\ell}$. $\zeta=13$ and $\alpha/\kappa\approx 1$. Again we also include a plot of where ${\cal R}_{\rm M}=1$ in the right-hand panel of Fig.~\ref{fig:fl0}, but this time the value of ${\tilde\Gamma}{\rm a}$ as a function of $p$ for different values of $m_*$.

In Fig.~\ref{fig:varyfl} we present $m_{\rm a}$ and $f_{\rm obs}$ as a function of ${\cal F}_\ell$ for various $q=p$ and $m_*=n_*=$  fixing $\zeta=13$, $\langle v^2\rangle=0.35$, ${\hat \Gamma}_{\rm a}=46$ and ${\tilde\Gamma}_{\rm a}=500$ while varying $\alpha/\kappa$ and $q$. For $\alpha/\kappa=0.1$ and 1 we see that $m_{\rm a}$ decreases as ${\cal F}_\ell$ increases, but the opposite is true for $\alpha/\kappa=10$ due to the relative sizes of the loop and long string contributions. It is clear that there is some value of  $1<\alpha/\kappa<10$ where contributions from loops and long strings are the same and one would get a flat line.  

\begin{figure}
    \centering
    \includegraphics[trim={0.1cm 0cm 1.5cm 1.3cm},clip,width=1\linewidth]{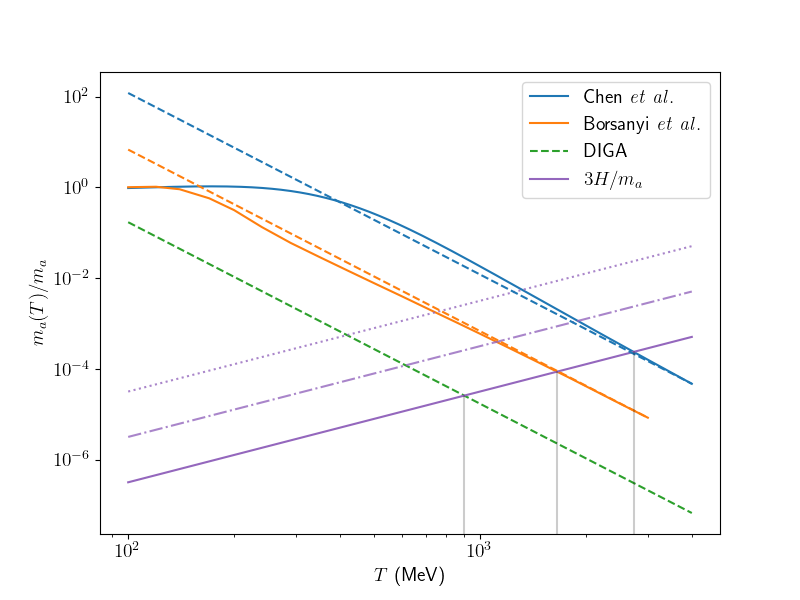}
    \caption{A comparison of predicted mass ratios $m_a(T)/m_a$ to $3H/m_a$ using $m_a = \{1,\, 10\,,100\}\mu{\rm eV}$ from top to bottom, with the intersection points corresponding to the inferred value of $T_{\rm AD}$ for each approach. For the Chen {\it et al.} results \cite{PhysRevD.106.074501} we use the fitting formula that they provide (note that it is extended to higher temperatures as the data that the fit was based upon only reaches up to 512 MeV) while the Borsanyi {\it et al.} plot uses the direct numerical results that are provided in Table S9 of the Supplementary Information \cite{Borsanyi:2016ksw}. Both works calculate the topological susceptibility, which we have converted to mass ratios using the value for $\chi(0)$ provided in \cite{Borsanyi:2016ksw}. Uncertainties on these calculations are provided in the original works, but we neglect to show them here in order to more easily compare the results to the power law scaling $T^{-4}$ (shown as dashed lines) and also because, for our purposes, the dominant uncertainty is the disagreement between different approaches to lattice QCD calculations and also with the DIGA. }
    \label{fig: T_AD plot}
\end{figure}

Finally, we point out other possible uncertainties our calculations not related to the evolution of the string network. The measurement of $\Omega_{\rm a}h^2$ by the {\it Planck} Satellite has sub-percent uncertainty and, within the Standard Model, the evolution of ${\cal N}$ using the QCD epoch is relatively well understood. Uncertainties from these two sources can only be significant if either the axions are not the entire dark matter density, or if there is other Beyond Standard Model physics. The key uncertainty that we wish to highlight comes from the power law prediction in equation (\ref{eq: DIGA mass}) that is predicted by the dilute instanton gas approximation (DIGA), and how it compares to lattice QCD calculations of the topological susceptibility. Most lattice QCD studies find $m_a(T) \propto T^{-4}$ to be a reasonable description in the high temperature limit, but there remain significant disagreements between different groups, and with the DIGA, that result in significantly different estimates for $T_{\rm AD}$. In Fig~\ref{fig: T_AD plot}, we display the mass ratio $m_a(T)/m_a = (\chi(T)/\chi(0))^{1/2}$, where $\chi$ is the topological susceptibility, using the results from two different groups \cite{PhysRevD.106.074501,Borsanyi:2016ksw} as well as from equation (\ref{eq: DIGA mass}). Other groups \cite{Bonati:2015vqz, PETRECZKY2016498} have also performed this calculation, but we do not present them here --- see \cite{PhysRevD.106.074501} for a more detailed summary plot. All of the dotted lines represent the $T^{-4}$ power law scaling, demonstrating that the three approaches broadly agree on the scaling with temperature, but disagree by orders of magnitude on the overall amplitude.

The inferred value for $T_{\rm AD}$ is given by the points of intersection with $3H/m_a$, which we have displayed for $m_a=1\mu{\rm eV}$, $10\mu{\rm eV}$ and $100\mu\rm{eV}$, showing that $T_{\rm AD}$ varies by a factor of $\sim 3$ in each case. For reasonable axion masses ($\gtrsim 10^{-2} \mu{\rm eV}$) all three methods for predicting the temperature dependent axion mass are well described by the form in equation \eqref{eq: DIGA mass}, with broad agreement on both $\bar{T}$ and $\beta_2$, but large variations across almost $3$ orders of magnitude in $\beta_1$. The most relevant uncertainty can, therefore, be parameterised as
%
%%%%%%%%%%%%%%%%%%%%%
\begin{equation}
    m_a\propto \beta_1^{-{1\over 3+\beta_2}}\,.
\end{equation}
%%%%%%%%%%%%%%%%%%%%%
%
with $\beta_2 \approx 4$ such that the axion mass depends upon $\beta_1^{-1/7}$. If the results of \cite{PhysRevD.106.074501} were to be preferred over the DIGA estimate that we have used, the predicted axion mass would be smaller by a factor of $\sim 2.6$.

Unfortunately, the difficulties in reliably calculating the topological susceptibility mean that even if we could perfectly model all of the physics relating to the production of axions in the Early Universe, we would still suffer from large uncertainties on the axion mass due to the lack of a consensus on QCD physics.

\section{Discussion and conclusions}
\label{sec:discussion}

Let us first summarise the key points that we have made in this paper, before discussing each in more detail:

\begin{itemize}
    \item The relic density of axions is very sensitive to the spectrum of radiation emitted by global strings. We have also argued that the contribution from all sources can be separated out into the product of a universal term that arises from cosmology and another term that depends on the details of each source, including the spectrum of radiation they emit. Further simulations of both networks of global strings and hybrid networks of strings and domain walls will be necessary to determine whether it is the IMM, strings or domain walls that provide the dominant contribution to the axion density, as it is independent of cosmology and typically determined by the emission spectrum.
    \item For a single oscillating global string, the propagating axion radiation needs to be carefully decoupled from the localised self-field of the string. If this is not accounted for, the resulting spectrum will be dominated by the motion of the string, which can drastically change the features of the spectrum and the inferred relic density.
    \item State of the art network simulations do not appear to have sufficiently accounted for this effect and are likely extracting a spectral index which does not accurately reflect the propagating radiation. The extent to which it needs to be corrected is yet to be seen in a network context, but it is noteworthy that the correction is significant for an oscillating string.
    \item We have shown how the axion mass varies with the spectral index in Fig~\ref{fig:fl1} and Fig~\ref{fig:fl0}, representing scenario A and B respectively, assuming that $\Omega_{\rm a}h^2 \approx 0.12$ and under reasonable choices for the other network parameters. In the former case, taking the limit $q\to\infty$ and $\alpha/\kappa \approx 1$ would be compatible with Nambu string simulations and leads to $m_{\rm a}\approx 160\,\mu {\rm eV}$ and $f\approx 38\,{\rm GHz}$. The latter represents the results from most modern field theory simulations of string networks without any extrapolation if $p\approx 1$. For spectral indices around and below 1, the cut-off scale becomes important for determining the relic density; setting $\tilde\Gamma_a = 500$ and $m_* \approx e^{70}$ leads to $m_a \approx 4\,\mu {\rm eV}$ and $f \approx 1\,{\rm GHz}$.
\end{itemize}

The sensitivity to the spectrum of radiation has been well understood for some time - hence scenarios A and B - but in section 2 we have delineated this issue into the function $G_2$ which comes about due to the integral required to calculate $n_{\rm a}$ from the spectral density. We have introduced the fraction of the scaling density that goes into loops and is directly emitted by the long string networks as a parameter, ${\cal F}_\ell$. When there is a significant fraction of the energy which goes into loops the energy is reprocessed to different frequencies governed by the function $G_1(\alpha/\kappa)$ which we have now calculated exactly, rather than using the approximation of previous work which ignored the effects of the spectrum --- that is, it effectively had $G_2=1$ since it worked in the regime where $q\gg 1$ and $n_*\gg 1$. The energy emitted by long strings is parameterized by the scale $\gamma$ which is difficult to estimate. Ultimately, we have shown that the relic density can be written as $\Omega_{\rm a}\approx N_{i}\Gamma_{\rm a}\times ..\times G_{2}$ where $N_i$ is an appropriate normalization factor with the same dependence of physical parameters, $\Gamma_{\rm a}=\left\{{\hat\Gamma}_{\rm a},{\tilde\Gamma}_{\rm a},\hat{\Gamma}_{\rm a, dw}\right\}$ and there can be additional suppression factors, such as $G_1$ in the case of loops. However, within the modeling, one might argue that there is significant uncertainty.

One might expect that simulations of the network evolution can give us a handle on some of these parameters. While we certainly believe this to be true, in many cases this needs to be done with great care. We have calculated the spectrum of radiation from a single string and explained how one has to make sure that the self-field of the string has been removed, which is not always trivial. It became clear that if this was not done properly, the signal would be dominated by the motion of the string and not by genuine propagating axions; an effect which can be seen directly in Figs.~\ref{string_cross_section_phis_2} and \ref{circular_mask_field}. In our simulations, this led to a markedly different spectrum, one which we have shown to agree well with the expectations from the Nambu-Goto action --- the total power follows $P\propto\varepsilon^4$ and the power emitted in each mode goes like $P_n\propto e^{-rn}$ with $r\approx 2.5 - 2.9$ --- albeit with a noticeable relaxation time. If we were to take this seriously and consider axion emission that is dominated by long strings, similar to scenario B, but instead using the instantaneous emission spectrum found in our simulations, then we would infer a mass of $m_a \approx 125\mu{\rm eV}$, which is substantially larger than the $m_a \approx 4\mu{\rm eV}$ inferred for a power law spectrum with $p=1$.

Of course, a simple oscillating string is a much more controlled scenario than a network of strings, and we are not suggesting that the spectra will necessarily be similar, we use it simply to demonstrate the sensitivity of the mass to the spectral index. The key point is that if a given method for extracting the spectrum does not work well in such a simple scenario, then there is no reason to expect it to work well in the more complicated case. This is concerning considering that the approach taken by the majority of large-scale simulations is exactly the approach that performs poorly here. It is not clear how easily the alternative methods used in this work can be applied to network simulations, as the self-field subtraction is made significantly harder by the complex geometry of the network and masking may require that a significant fraction of the simulation volume is removed, much larger than has been attempted previously. 

Another technique for suppressing the self-field contribution was recently proposed in~\cite{Correia:2025nns}, where the contribution to a new quantity $J_{\rm{s}}$, formed by taking a radial projection of the PQ current in momentum space, was shown to scale as a string velocity squared. That is one order of $v$ more than usually computed $\phi\partial_{t}\alpha$. Investigating other projections, they claimed that the self-field constitutes $\approx30\%$ of the spectrum of $\phi\partial_{t}\alpha$ in a network simulation. This result provides further evidence to support our concern about the accuracy of the axion spectrum estimations available in the literature. As such, we do not advocate for any specific technique, in fact we emphasise that further advancements in isolating the axions from the self-field would be very useful. However, we suggest that simple simulations, where it is known how to remove the self-field to a large extent, should be used as a verification stage, before they are applied to large-scale simulations.

\section*{Acknowledgements}
We are grateful for the discussions with the participants of the \enquote{Topological defects in Cosmology} workshop held at the University of Manchester and \enquote{Early Universe from Home 2026} conference. We want to thank Mark Hindmarsh, Jose Correia and Amelia Drew for helpful comments and discussion of this work. In addition, we would like to acknowledge  collaboration with Shikhar Agarwal on early parts of this work and helpful comments from Oleksandr Shelestiuk. The work of LB is supported by the STFC Doctoral Training Award No. ST/X508597/1. The simulations presented in this work were conducted on the resources of the Computational Shared Facility at the University of Manchester.

\bibliography{double_column}

\appendix
\section {Tests of the calculations of the power for $n_x=200$ and $400$}
\label{sec:small_boxes}
\begin{figure*}
    \centering
\includegraphics[scale=1]{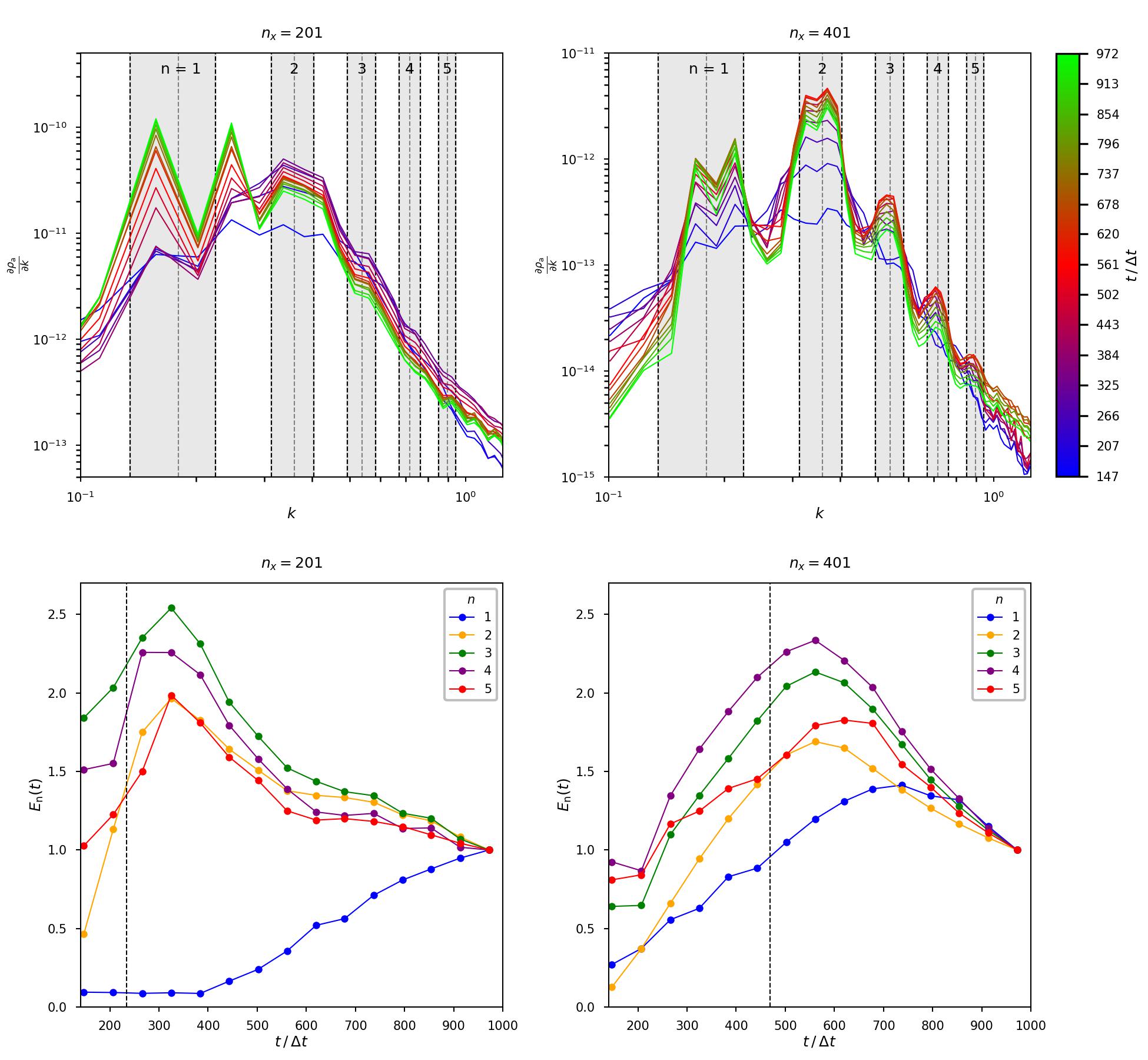}

    \caption{In the top row we present the evolution of the spectrum for the simulations with $n_{x} = 201$ (on the left) and $401$ (on the right) and $L = 50\,\Delta x$. In the bottom row we present the energy evolution for each peak. The vertical dashed line shows the time when the radiation hits the boundary of the box.}    
    \label{201_401_calc}
\end{figure*}
\label{sec:power_calc_appendix}
To investigate the influence of the simulations’ finite volume on the behavior, we observe, we compare results obtained from the simulations with different box sizes $n_{x} = 201$ and $401$. The parameters of these simulations, for example grid spacing, size of the time step, and the length of the simulation were kept the same as in section \ref{sec:straight}. To simulate the same physical scenario, we kept $\varepsilon$ and the wavelength of the string’s oscillation the same.

To search for the difference introduced by varying box size, we show the evolution of the energy spectra for both simulations in the top row of Fig. (\ref{201_401_calc}) together with the evolution of the energy stored in each peak identified in section \ref{sec:straight} in the bottom row. Unlike the spectra for the simulation with $n_x = 801$ the spectra shown in this figure were not rebinned and are plotted with the maximal resolution we can obtain for the momentum for these simulations, as detailed in appendix~\ref{sec:reconstruction_alg}. Both simulations show spectra with an overall shape similar to the spectra shown in Fig.~\ref{spectrum_evolution}, but differ significantly in details. The spectra for $n_x =201$ simulation show no feature that could be interpreted as the peaks centered at the harmonics of the string. For the simulation with bigger $n_x$ , the energy spectra obtained from the simulation show clearly the formation of peaks, whose height, initially increasing, starts decreasing at later times. 

This behavior can be clearly seen in the plots presented in the bottom row of the figure, that show the integral of the spectrum over a band of width 0.09 centered around each peak. The energy for each peak, obtained in this way, is plotted as a function of time in the bottom row of the figure, together with vertical dashed lines showing the moment when we expect the radiation to hit the boundary of the box calculated using $t_{\rm{b}} = n_{x}dx/(2dt)$. The plots clearly show a significant change in the behavior of the peaks’ energy soon after this time. The initially growing contributions to the n=2,3,4,5 modes start decreasing due to absorption of the radiation by the boundary. At this time the boundary starts absorbing the radiation hitting it, causing the height of the peaks to decrease. We also note that due to numerical imperfections, the boundary reflects a small fraction  of the radiation back into the volume of the simulation. The reflected radiation can then interfere with the rest of the field disturbing the formation of the spectrum and the behavior of the string.  

As the rate at which the boundary absorbs the radiation is unknown, we cannot perform any analysis for times later than $t_{\rm{b}}$ and need to constrain the simulation to end before radiation hits the boundary. Therefore, the $n_{x}=801$ simulations used in the main analysis were ran until $t\approx1000\,\Delta t$, that is within one period of the string's oscillation after $t_{\rm{b}}$. The time taken for the peaks in the axion energy spectrum to form properly, is much greater than $t_{\rm{b}}$ for simulations with small box sizes e.g. $n_{x} = 101, 201, 401$ \cite{battye1994global} and requires simulations with much larger volumes. We therefore conclude that the best results can only be obtained by analyzing the simulation with $n_{x }= 801$.

\section {Derivation of the mode-mixing kernel and the reconstructed spectrum}
\label{sec:reconstruction_alg}

In this appendix, we present a derivation of the mode-mixing kernel that is an adaptation of the computation given in appendix~A1 of \cite{Hivon:2001jp} to 3-dimensional space.

In the derivation, we used the Fourier transforms defined with non-unitary (so-called "physics") normalization convention and denoted by $\widehat{}$. We also use $\bar{}$ to denote masked quantities.

The mode mixing kernel $K_{\rm{k'k}}$ can be defined by the relation 
\begin{equation}
\begin{split}
   \widehat{\overline{\partial_{t}\alpha}} = \int (\partial_{t}\alpha) W(\mathbf{r}) e^{-i\mathbf{k}\cdot\mathbf{r}}d^{3}\mathbf{r} \\
   \equiv \int \widehat{(\partial_{t}\alpha)}(\mathbf{k}) K_{k'k}[W] d^{3}\mathbf{k'},
\end{split}
\end{equation}
in terms of the masking function $W(\mathbf{r})$. By insertion of the inverse Fourier transform of $\widehat{(\partial_{t}\alpha)}(\mathbf{k'})$ into the middle expression and equating it to the integral on the right, we express the kernel $K_{k'k}$ as
\begin{equation}
    K_{k'k}[W] = \frac{1}{(2\pi)^{3}} \int e^{i\mathbf{k'}\cdot \mathbf{r}} W(\mathbf{r})e^{-i\mathbf{k}\cdot \mathbf{r}}d^{3}\mathbf{r}. 
\end{equation}
This expression can be further simplified using the inverse Fourier transform of the $\widehat{W}(\mathbf{k''})$. Collecting exponentials and noting that the only dependence on $\mathbf{r}$ enters through the argument of the exponential function, we can perform the integral over $\mathbf{r}$ resulting in
\begin{equation}
\label{App_3}
    K_{k'k}[W] = \frac{1}{(2\pi)^{3}} \int \widehat{W(\mathbf{k})} \delta(\mathbf{k''}+\mathbf{k'}-\mathbf{k})d^{3}\mathbf{k''}. 
\end{equation}
We continue by defining the ensemble average of the masked spectrum as
\begin{equation}
    \langle \overline{P}(k) \rangle = \frac{k^{2}}{(2\pi L)^{3}} \int d\Omega_{k} \langle \widehat{\overline{(\partial_{t}\alpha)}}^{*}(\mathbf{k}) \widehat{\overline{(\partial_{t}\alpha)}}(\mathbf{k})\rangle. 
\end{equation}
The masked spectrum can be expressed as
\begin{equation}
\begin{split}
\label{App_5}
    \langle \overline{P}(k) \rangle = \frac{k^2}{(2\pi L)^3} \int d\Omega_{k} \int d^{3}\mathbf{k'} \int d^{3}\mathbf{k''} K_{k'k}^{*}K_{k''k} \\
    \langle \widehat{(\partial_{t}\alpha)}(\mathbf{k'})^{*}\widehat{(\partial_{t}\alpha)}(\mathbf{k''})\rangle.
\end{split}
\end{equation}
Using (\ref{App_3}) and the relation
\begin{equation}
    \langle \widehat{(\partial_{t}\alpha)}(\mathbf{k'})^{*}\widehat{(\partial_{t}\alpha)}(\mathbf{k''})\rangle = \frac{(2 \pi)^{5}}{k'^{2}} \delta^{(3)}(\mathbf{k'}-\mathbf{k''})\langle P(k')\rangle,
\end{equation}
we rewrite (\ref{App_5}) as
\begin{equation}
\begin{split}
    \langle \overline{P}(k) \rangle = \frac{k^2}{2\pi(2\pi L)^3} \int dk' \langle P(k')\rangle\int d\Omega_{k'}\int d\Omega_{k} \\    
      \int d^{3}\mathbf{k''} \widehat{W}(\mathbf{k''})^{*} \widehat{W}(\mathbf{k''}) \delta^{(3)}(\mathbf{k'}-\mathbf{k}+\mathbf{k''}).
\end{split}
\end{equation}
The above expression can be simplified to
\begin{equation}
\label{App_8}
    \langle \overline{P}(k) \rangle = \frac{k^2}{(\pi)^{2}} \int dk' \langle P(k')\rangle \int d k'' P_{W}(k'') J(k,k',k''),
\end{equation}
where we defined the spectrum of the masking function as
\begin{equation}
    P_{W}(k) = \frac{k^2}{(2\pi L)^{3}} \int d\Omega_{k} \lvert \widehat{W}(\mathbf{k)}\rvert^{2},
\end{equation}
and 
\begin{equation}
    J(k,k',k'') = \frac{(2\pi)^3}{(4\pi)^{2}} \int d\Omega_{k'}d\Omega_{k} \delta^{(3)}(\mathbf{k'}-\mathbf{k}+\mathbf{k''}),
\end{equation}
which takes a value
\begin{equation}
    J(k,k',k'') = \frac{(\pi)^{2}}{kk'k''},
\end{equation}
in the interval $L: \lvert k'-k'' \rvert \leq k \leq k' + k''$ and is 0 otherwise.

Expression (\ref{App_8}) can be rewritten as 
\begin{equation}
    \frac{\langle \overline{P}(k) \rangle}{k} = \int dk' \left( \int_{L} dk'' \frac{P_{W}(k'')}{k''}\right)\frac{\langle P(k')\rangle}{k'} ,
\end{equation}
and cast into the following matrix form
\begin{equation}
    \frac{\langle \overline{P}(k) \rangle}{k} = \int dk' \mathcal{M}^{-1}(k,k')\frac{\langle P(k')\rangle}{k'}.
\end{equation}
We use the following relation
\begin{equation}
    \int dk \mathcal{M}(k'',k)\mathcal{M}^{-1}(k,k') = \delta(k''-k'),
\end{equation}
to invert the matrix $\mathcal{M}^{-1}(k,k')$. As a result, we get the equation for the ensemble averaged reconstructed spectrum, given by
\begin{equation}
    \frac{\langle P(k'')\rangle}{{k''}} = \int dk \mathcal{M}(k'',k)\frac{\langle \overline{P}(k) \rangle}{k}.
    \label{app_15}
\end{equation}
We note that the equivalent expression was presented in \cite{saikawa2024spectrum, Kim:2024wku} and a similar approach was developed in \cite{PhysRevD.83.123531}.\\

In the analysis presented in the main body of the paper, we use the above relation without the ensemble averages. This approximation assumes that the average performed on the set of vectors with $\lvert \mathbf{k} \rvert = k$ (or $k'$ appropriately), as part of the spectrum's computation, has converged to its ensemble average. In the discretized momentum space that approximation only holds for the k-bins with sufficiently large momenta. This means that the reconstruction of the spectra may fail for low values of the momentum. This obstacle poses no problem to the analysis performed in the main body of the paper, as we do not use the information from the first few bins of the spectra when extracting information from them.\\

For the simulation performed on a cubic lattice, the spectra of $q = \{ \partial_{t}\alpha(\mathbf{r}), \overline{\partial_{t}\alpha}(\mathbf{r}), W(\mathbf{r}) \}$are calculated using 
\begin{equation}
     P_{q}(k') = \sum_{k = k'-\frac{\Delta k}{2}}^{k = k'+\frac{\Delta k}{2}}\left( \frac{(d k)^{3}}{\Delta k (2\pi L)^{3}}\right) \lvert \widehat{q}(\mathbf{k})\rvert^{2},       
\end{equation}
where $q(\mathbf{k})$ is a discrete Fourier transform of $q(\mathbf{r})$, that is performed using FFTW 3.3.10 \cite{FFTW05} package. The above expression is obtained by approximating the solid angle element $d\Omega_{k}$ by the ratio of the volume element in momentum space and the width of the bin $\Delta k$, which can be seen as approximating the solid angle integral by the volume integral over a thin shell divided by the width of the shell. This approximation holds for the bins with a large momentum, but breaks down as $k\rightarrow0$. As mentioned before we do not extract the information using the bins with low momenta, so our results are not affected by this approximation, except for the case when we use the spectrum of the mask to construct the correction matrix. In this case the first bin of the spectrum contains information about the reduction of the volume containing information due to masking. To properly capture this information in the spectrum of the mask we set the number of bins in the spectrum to be exactly equal to $n_x/2$. We than note that thanks to that the first element of the spectrum contains only the contribution from the point $(0,0,0)$, which in turn is nothing else than an average of the mask. We proceed by setting the first element of the spectrum of the mask to be equal to the average of the mask (evaluated before taking the Fourier transform). 

The spectrum of the mask is than used to evaluate the matrix in \ref{app_15}) by approximating the integral as 
\begin{equation}
     \mathcal{M}^{-1}(k,k') = \sum_{\lvert k'-k''\rvert}^{k'+k''}  \frac{P_{W}(k'')}{k''}\,,
\end{equation}
where $k,k',k''$ are taken to be central values of each bin of the spectrum. The matrix $\mathcal{M}^{-1}(k,k')$ is inverted and used to evaluate the reconstructed spectrum $P(k'')$ using
\begin{equation}
        \frac{ P(k'')}{{k''}} = \sum_{k=0}^{k = \frac{\pi}{2}} \mathcal{M}(k'',k)\frac{\overline{P}(k)}{k} \,,
\end{equation}
where $\overline{P}(k)$ is the masked spectrum. 

\end{document}